\documentclass[journal]{IEEEtran}
\usepackage{xcolor}
\newcommand{\algcomment}[1]{\hfill{\color{gray}\textit{\scriptsize // #1}}}

\usepackage{cite}
\usepackage{paralist}
\usepackage[cmex10]{amsmath}
\usepackage{amssymb, amsfonts}
\usepackage{graphicx}
\usepackage{epstopdf}
\usepackage{balance}
\usepackage{stfloats}
\usepackage{array}
\usepackage{bm}
\usepackage{color}
\usepackage{amsmath}
\usepackage{blkarray}
\usepackage{makecell}
\usepackage{booktabs}
\usepackage{soul}

\usepackage{mathrsfs}
\DeclareUnicodeCharacter{25C4}{\triangleleft}
\usepackage[caption=false,font=footnotesize]{subfig}
\IEEEoverridecommandlockouts  
\usepackage{url}             
\usepackage[
  paper=letterpaper,
  top=0.75in,
  bottom=1in,
  inner=0.625in,
  outer=0.65in,
  columnsep=0.22in
]{geometry}

\usepackage{algorithm,algorithmic}

\newtheorem{remark}{Remark}

\newcommand{\green}[1]{{\textcolor[rgb]{0.14,0.4,0.19}{#1}}}

\newcommand{\blue}[1]{{\textcolor[rgb]{0,0,1}{#1}}}

\newcommand{\red}[1]{{\textcolor[rgb]{1,0,0}{#1}}}

\newcommand{\gray}[1]{{\textcolor[rgb]{0.5,0.5,0.5}{#1}}}

\begin{document}

\title{Semantic Error Correction and Decoding\\ for Short Block Codes}

\author{Jiafu Hao, \IEEEmembership{Student Member, IEEE},
        Chentao Yue, \IEEEmembership{Member, IEEE},
        Wanchun Liu, \IEEEmembership{Member, IEEE},\\
        Branka Vucetic, \IEEEmembership{Fellow, IEEE}
        and Yonghui Li, \IEEEmembership{Fellow, IEEE},%
\thanks{Jiafu Hao, Chentao Yue Wanchun Liu, Branka Vucetic and Yonghui Li are with the School of Electrical and Computer Engineering, The University of Sydney, Sydney, NSW 2006, Australia
(e-mail: \{jiafu.hao; chentao.yue; wanchun.liu; branka.vucetic; yonghui.li\}@sydney.edu.au). (\textit{Corresponding author: Chentao Yue})}%

\thanks{Code available: https://github.com/Jeh100/Semantic-Error-Correction-and-Decoding. The work of Chentao Yue was supported by ARC under Grant DE250101332.}%
}

\IEEEaftertitletext{\vspace{-2\baselineskip}}
\setlength{\abovedisplayskip}{2pt}
\setlength{\belowdisplayskip}{2pt}

\maketitle
\thispagestyle{empty}

\begin{abstract}
This paper presents a semantic-enhanced receiver framework for transmitting natural language sentences over noisy wireless channels using multiple short block codes. After ASCII encoding, the sentence is divided into segments, each independently encoded with a short block code and transmitted over an AWGN channel. At the receiver, segments are decoded in parallel, followed by a semantic error correction (SEC) model, which reconstructs corrupted segments using language model context. We further propose the semantic list decoding (SLD), which generates multiple candidate reconstructions and selects the best one via weighted Hamming distance. Moreover, a semantic confidence-guided HARQ (SHARQ) mechanism is designed to replaces Cyclic Redundancy Check (CRC) with a semantic confidence score, enabling selective segment retransmission. We analyze the block error rate (BLER) for the proposed framework and discuss the tradeoff between the semantic gain from segmentation  and the finite-blocklength penalty of shorter codes. Simulation results demonstrate that SEC provides approximately 0.4 dB BLER gain over plain short-code transmission, while SLD extends this to 0.8 dB. Compared to transmitting the entire sentence as a single long 5G LDPC codeword, our approach significantly improves semantic fidelity and reduces decoding latency by up to 90\%. SHARQ further provides an additional 1.5 dB gain over conventional HARQ.
\end{abstract}

\begin{IEEEkeywords}
Short block codes, semantic error control, large language models, hybrid automatic repeat request.
\end{IEEEkeywords}

\vspace{-0.5em}
\section{Introduction}
\vspace{-0.3em}
\IEEEPARstart{U}{ltra}-reliable and low-latency communications (URLLC) is one of the key 5G service paradigms. The design of the physical layer, especially the channel coding scheme, for URLLC involves a fundamental trade-off between latency and reliability \cite{yue2023efficient}. Long blocklength codes, such as LDPC codes \cite{1057683}, can approach the Shannon capacity under Belief Propagation (BP) \cite{Mahyar2019ShortCode}. However, their inherent long blocklength introduces significant propagation and processing latency, making them unsuited for URLLC. This motivates the use of short blocklength codes to meet the tight latency requirements. However, as established by the finite blocklength theory \cite{PPV}, shorter blocklengths inevitably lead to a degradation in error-correction capability, posing a fundamental challenge for achieving ultra-reliable transmission.

Semantic communication (SemCom) has emerged as a promising paradigm for future communication systems, shifting the focus from bit-level transmission to the extraction and delivery of meaning-relevant information \cite{9955525}. Unlike conventional approaches that prioritize every bit accuracy, SemCom aims to preserve the fidelity of transmitted meaning. Current SemCom research follows two primary directions. The first is source–channel separation coding design. They only focus on source coding, aiming to compress and transmit essential semantic content efficiently. For example, \cite{10494374} introduced importance-weighted semantic triples to identify key semantic information, while \cite{Lee} employed VQ-VAE for semantic compression to reduce transmission volume. \emph{Liu et al.} \cite{LIU} exploited context information within and between sentences for enhanced semantic representation and recovery. The second direction is joint source-channel coding (JSCC) that optimizes encoders and decoders jointly through end-to-end training \cite{jscc}. JSCC completely replaced the traditional structure with a neural network. DeepJSSC \cite{deepsc} presents a deep neural network framework that integrates semantic coding and channel coding for end-to-end transmission. Swin-JSCC \cite{swin-jscc} leveraged the Swin Transformer architecture and adaptive modules to achieve high-performance. D²-JSCC \cite{d2jscc} combined deep source coding with adaptive density models and digital channel block coding to minimize end-to-end distortion. While these JSCC approaches achieve strong performance by jointly optimizing the entire transmission chain, they require a complete replacement of conventional communication modules with neural networks, violating the source--channel separation principle~\cite{6773024} and incurring high deployment costs for practical communication systems.

Recent advances have explored integration between SemCom and physical layer design. Some works focus on system-level integration, where \emph{Lee et al.} \cite{Lee} integrated pre-trained language models with 5G-NR physical layer functions, and Evgenidis et al. \cite{hybird_semcom} proposed a hybrid semantic-Shannon multi-carrier system that jointly optimizes transmission mode selection and power allocation. Other works exploited semantic information to enhance specific physical-layer functionalities. For example, \cite{MIMO1} jointly designed semantic coding and massive MIMO beamforming, \cite{LLM-aid} proposed a semantic pilot scheme leveraging LLM-corrected text for data-aided channel estimation, and \cite{TWRC} developed a semantic-empowered physical-layer network coding framework for two-way relay channels. Semantic information has also been leveraged to improve transmission reliability and security, including semantic signals as information-bearing artificial noise for physical-layer security \cite{secrecy}.
Semantic-aware hybrid automatic repeat request (HARQ) has been explored for vehicular networks~\cite{harq}, but it operates within an end-to-end JSCC framework and lacks fine-grained segment-level retransmission guided by channel soft information.

Despite these advances, SemCom faces critical challenges for practical deployment in latency-sensitive applications. Source--channel separation SemCom systems typically require long codewords to represent semantic content~\cite{9955525}, introducing transmission delays that conflict with strict latency requirements. These limitations motivate a different approach: rather than redesigning the entire communication chain, semantic information can be selectively incorporated into the physical layer, preserving compatibility with existing infrastructure. In this direction, Kim~et~al.~\cite{bit_flip} proposed unequal error protection for digital semantic communication, where learned bit-flip probabilities were used to allocate coding redundancy according to semantic importance. Park and Yang~\cite{LLM-aid} leveraged LLM-corrected text to identify reliable decoded symbols as semantic pilots for data-aided channel estimation. However, \cite{bit_flip} focused on encoder-side redundancy allocation without exploiting semantic information at the decoder, while \cite{LLM-aid} operated on uncoded symbols and does not integrate with channel coding. The potential of semantic information to directly improve channel decoding performance remains unexplored.

To address these issues, we propose a receiver-side framework that integrates a pretrained language model into the decoding process of standard short block codes, preserving the conventional source--channel separation architecture. We focus on natural-language text in this work, as it is the dominant payload in machine-to-machine signaling and the short control messages in URLLC and IoT applications.  Rather than transmitting a natural-language sentence as a single long codeword, the sentence is partitioned into short segments, each independently channel-encoded and decoded. Segments that fail channel decoding are then reconstructed through semantic inference conditioned on the correctly decoded segments. The key observation is that a single long codeword, if it fails during decoding, results in the loss of all contextual information. In contrast, dividing the message into multiple short codewords confines decoding errors to specific segments and leaves the remaining segments intact, providing the language model with reliable context for reconstruction. While this work focuses on text, the principle applies to any structured source. The main contributions of this paper are summarized as follows.

\begin{itemize}
    \item We propose a multiple short-code (MSC) transmission framework, in which a natural-language sentence is partitioned into segments, each independently encoded by a short block code. We show that localized channel-decoding failures preserve the semantic context. Building on MSC, we propose a semantic error correction (SEC) module that replaces erroneous segments with semantically coherent reconstructions generated by a fine-tuned bidirectional and auto-regressive transformers (BART) model, exploiting cross-segment context through bidirectional self-attention.
    \item To overcome the unverifiable nature of single-output semantic correction, that semantically plausible but incorrect reconstructions cannot be distinguished from correct ones without channel information. We propose semantic list decoding (SLD). For each unreliable segment, SLD generates a list of candidate reconstructions via diverse beam search, re-encodes each candidate, and selects the one minimizing the weighted Hamming distance against the received soft information. This approach couples language-model reasoning with conventional bit-level channel reliability.
    \item We propose a semantic confidence-guided HARQ (SHARQ) scheme, which replaces CRC-based error detection with a soft-information-based confidence metric. SHARQ eliminates the CRC overhead that is prohibitive for short block codes and enables segment-selective retransmission.
    \item We derive closed-form block error rate (BLER) expressions for MSC with semantic processing. Via conditional entropy and Fano's inequality, we provide an information-theoretic interpretation that quantifies  the tradeoff between the semantic gain from segmentation and the finite-blocklength penalty of shorter codes.
\end{itemize}

Simulation results on the SNLI corpus over the binary-input AWGN channel confirm the effectiveness of the proposed scheme.  Under single-shot transmission, the SEC and SLD pipeline achieves a $1.1$ dB BLER gain at $0$ dB SNR over a single $(1024,512)$ 5G LDPC codeword while maintaining BLEU $> 93$ and ROUGE-L $> 97$. SLD recovers $99\%$ of erroneous $(32,16)$ segments, and parallel short-code decoding reduces per-sentence latency by $76$--$90\%$ relative to long-LDPC decoding. When the retransmission is enabled, SHARQ yields an additional $1.5$ dB gain over CRC-based HARQ at the same target BLER.

The remainder of this paper is organized as follows. Section~II reviews the background of our proposed methods. Section~III presented the MSC framework and the SEC module. Section~IV develops the SLD scheme, including error identification, candidate generation, and WHD-based selection. Section~V introduces the SHARQ retransmission mechanism. Section~VI provides the performance and complexity analysis. Simulation results and discussions are presented in Section~VII. Finally, Section~VIII concludes the paper.

\vspace{-1em}
\section{Background}
\label{section2}
\vspace{-0.3em}

\subsection{Linear block code} 
\label{section2a}
A binary linear block code $\mathcal{C}(n, k)$ encodes $k$ information bits into an $n$-bit codeword ($n > k$), where the $n-k$ redundant bits provide error-correction capability. The code is defined by its generator matrix $\mathbf{G} \in \{0,1\}^{k \times n}$, and encoding maps a message vector $\mathbf{b} \in \{0,1\}^k$ to a codeword $\mathbf{c} = \mathbf{b}\mathbf{G} \in \{0,1\}^n$. The codeword $\mathbf{c}$ is modulated via BPSK as $\mathbf{x} = 1 - 2\mathbf{c} \in \{-1,+1\}^n$ and transmitted over an additive white Gaussian noise AWGN channel. The received signal is
\begin{equation}
\mathbf{y} = \mathbf{x} + \mathbf{z},
\end{equation}
where $\mathbf{z} \sim \mathcal{N}(\mathbf{0}, \sigma^2 \mathbf{I}_n)$ is the i.i.d. Gaussian noise. The signal-to-noise ratio (SNR) is given by $\frac{1}{\sigma^2}$.

At the receiver, a decoder estimates the transmitted codeword as $\hat{\mathbf{c}}$. A decoding error occurs if $\hat{\mathbf{c}} \neq \mathbf{c}$. 

The best achievable BLER performance of block codes is fundamentally constrained by their blocklength. According to the normal approximation (NA) bound of the finite blocklength theory \cite{PPV}, for code $\mathcal{C}(n, k)$, its best BLER in binary AWGN channels is approximately given by \cite{erseghe2016coding}:
\begin{equation} \label{equ::PPV}
   \epsilon^*(k, n) \approx  Q\left( \sqrt{\frac{n}{V}}\cdot\left(\frac{C-R}{\log_2e}+\frac{\log n}{2n}\right)\right),
\end{equation}
where $C$ is the channel capacity, $V$ is the channel dispersion \cite[Fig. 6]{erseghe2016coding}, and $Q(\cdot)$ is the Gaussian Q-function. As shown by \eqref{equ::PPV}, at the same code rate $R$, the BLER degrades significantly as $n$ decreases, establishing the performance penalty of short codes.

\vspace{-0.3em}
\subsection{Ordered Statistics Decoding and Re-encoding}
\label{section2b}
Ordered statistics decoding (OSD)~\cite{554278} is a near-ML decoder applicable to any linear block code, operating directly on the generator matrix without requiring code-specific structure. The decoding complexity of OSD is roughly $\mathcal{O}(k^{m+2})$, where $m$ is the decoding order. For codes with minimum Hamming distance $d_{\min}$, order $m = \lceil d_{\min}/4 - 1 \rceil$ approaches ML performance~\cite{pbosd}.

\subsubsection{OSD Preprocessing}

We define $\alpha_i = |y_i|$ to represent the reliability of the received symbol $y_i$. The hard-decision vector $\mathbf{r} = [r]_1^n$ is obtained by thresholding: $r_i = 1$ for $y_i < 0$ and $r_i = 0$ otherwise. Prior to decoding, OSD performs two permutations on the received vector $\mathbf{y}$ and the generator matrix $\mathbf{G}$. The first permutation $\pi_1$ reorders $\mathbf{y}$ and the columns of $\mathbf{G}$ according to the descending order of $\boldsymbol{\alpha} = [\alpha]_1^n$. Subsequently, Gaussian elimination (GE) is applied to the permuted matrix $\pi_1(\mathbf{G})$ to convert it into systematic form, i.e.,
\[
\tilde{\mathbf{G}} = [\mathbf{I}_k \ \tilde{\mathbf{P}}],
\]
where $\mathbf{I}_k$ denotes the $k \times k$ identity matrix and $\tilde{\mathbf{P}}$ represents the corresponding parity sub-matrix. During the elimination process, an additional permutation $\pi_2$ may be introduced to guarantee that the first $k$ columns are linearly independent. After permutation, the reordered received vector and generator matrix can be expressed as
\begin{equation}
\tilde{\mathbf{y}} = \pi_2(\pi_1(\mathbf{y})), \qquad
\tilde{\mathbf{G}} = \pi_2(\pi_1(\mathbf{G})),
\end{equation}
respectively. Similarly, the reliability sequence and the hard-decision vector are rearranged as
\begin{equation}
\tilde{\boldsymbol{\alpha}} = \pi_2(\pi_1(\boldsymbol{\alpha})),
\qquad
\tilde{\mathbf{r}} = \pi_2(\pi_1(\mathbf{r})).
\end{equation}

For any length-$n$ vector, subscripts $(\cdot)_{B}$ and 
$(\cdot)_{P}$ denote its first $k$ and remaining $n-k$ components, respectively.

\begin{figure}[t]
\centering
\includegraphics[width=0.49\textwidth]{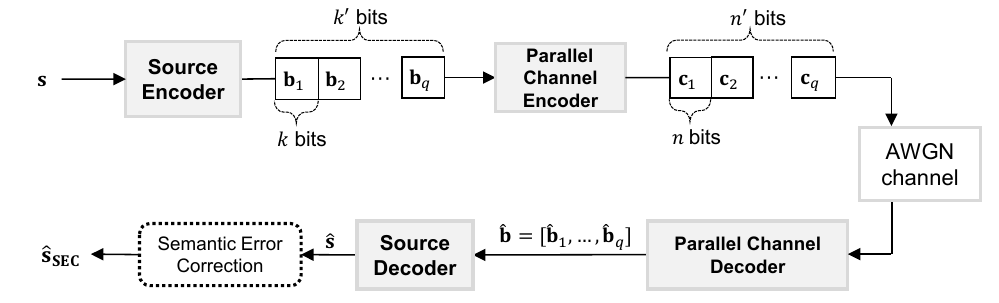}
\caption{Proposed MSC framework with parallel short block codes and SEC module. Modulation and demodulation are omitted for clarity.}
\vspace{-1.1em}
\label{fig1}
\end{figure}

\subsubsection{Re-encoding}
The key insight of OSD lies in the re-encoding step. Once $\tilde{\mathbf{G}}$ is in systematic form, any valid codeword is \emph{fully determined} by its $k$ systematic bits. Decoding therefore reduces to a search over hypotheses for the $k$-bit systematic part. Each hypothesis is expressed as a test error pattern (TEP) $\mathbf{e} \in \{0,1\}^k$ applied to $\tilde{\mathbf{r}}_{B}$, from which a complete candidate codeword is obtained by re-encoding:

\begin{equation}
\tilde{\mathbf{c}}_{\mathbf{e}}
= (\tilde{\mathbf{r}}_B \oplus \mathbf{e}) \tilde{\mathbf{G}}
= \left[
\tilde{\mathbf{r}}_B \oplus \mathbf{e}
\quad
(\tilde{\mathbf{r}}_B \oplus \mathbf{e}) \tilde{\mathbf{P}}
\right].
\end{equation}

TEPs are evaluated in ascending order of Hamming weight. The maximum allowed Hamming weight is referred to as the OSD order.
For BI-AWGN channels, the optimal codeword $\tilde{\mathbf{c}}_\mathrm{opt}$ minimizes the weighted Hamming distance (WHD):

\begin{equation} \label{eq:whd_osd}
d^{(W)}(\tilde{\mathbf{c}}_{\mathbf{e}}, \tilde{\mathbf{y}})
\triangleq
\sum_{\substack{1 \le i \le n \\ \tilde{c}_{\mathbf{e},i} \ne \tilde{y}_i}}
\tilde{\alpha}_i .
\end{equation}
Finally, the decoded output $\hat{\mathbf{c}}$ is obtained by applying the inverse permutations to $\tilde{\mathbf{c}}_\mathrm{opt}$, namely,
\begin{equation}
\hat{\mathbf{c}}
=
\pi_1^{-1}\big(
\pi_2^{-1}(\tilde{\mathbf{c}})
\big).
\end{equation}

The proposed SLD scheme exploits this re-encoding principle. Whereas OSD enumerates information candidates by applying TEPs, SLD obtains its candidates from a language model conditioned on the surrounding context. Each candidate text segment can be mapped to a $k$-bit information vector, re-encoded into a length-$n$ codeword, and ranked by the distance against the received signal. Re-encoding therefore connects semantic-level inference with bit-level channel reliability. The details will be introduced in Section \ref{section4}.

\vspace{-0.3em}
\subsection{Pretrained Language Models}
\label{section2c} 
The Transformer architecture \cite{NIPS2017_3f5ee243} captures contextual dependencies in sequential data through self-attention mechanisms. Building on this foundation, Bidirectional and Auto-Regressive Transformers (BART) \cite{bart} is a sequence-to-sequence model combining a bidirectional encoder with an autoregressive decoder (distinct from the channel encoder/decoder in Section \ref{section2a}).

BART is pre-trained as a denoising autoencoder. Specifically, given a corrupted input sequence (e.g., masking, deletion, and substitution), it learns to reconstruct the original text by minimizing a cross-entropy loss over the output tokens. This pre-training objective closely mirrors the channel decoding task, where channel decoding errors results in corrupted characters in the decoded text.

Architecturally, BART's encoder processes the entire input sequence via stacked Transformer layers \cite{NIPS2017_3f5ee243} to produce contextual representations, which capture dependencies across all input positions through self-attention. The decoder autoregressively generates output tokens conditioned on these representations and all previously generated tokens. For tokenization, BART employs byte-pair encoding (BPE) \cite{BPE} tokenization, which maps an input string into subword tokens drawn from a fixed learned vocabulary.

\section{Proposed Semantic Error Correction Scheme}
\label{section3}

\subsection{System Overview}
\label{section3a}

We consider the transmission of a natural language sentence $\mathbf{s}$ with character length $\ell$ over a noisy channel. Each character is first converted to its 8-bit ASCII representation, resulting in a binary bit stream $\mathbf{b}'$ with $k'=8\ell$ bits. 

\begin{table*}[t]
\centering
\caption{Examples of MSC with SEC for different error scenarios using the $(128,64)$ code at SNR = 2 dB.}
\renewcommand{\arraystretch}{1}
\begin{tabular}{@{} p{5.5cm} p{5.5cm} p{5.5cm} @{}}
\toprule
\textbf{Original Sentence $\mathbf{s}$} & \textbf{Before SEC $\hat{\mathbf{s}}$} & \textbf{After SEC $\hat{\mathbf{s}}_{\mathrm{SEC}}$} \\

\midrule
A lady in a blue shirt and hat dancing with \ldots\, & $\underbrace{\texttt{c/?ueK(y}}_{\text{Error}}$n a blue shirt and hat dancing with \ldots\ & \red{A girl i}n a blue shirt and hat dancing with  \\
\midrule
\ldots\,one little boy are running on the grass. & \ldots\,one little $\underbrace{\texttt{[?yX\#Gg?}}_{\text{Error}}$running on the grass. & \ldots\,one little \green{boy are} running on the grass. \\

\bottomrule
\end{tabular}
\label{tab1}
\vspace{-1.2em}
\end{table*}

\subsubsection{Multiple Short Code (MSC) Transmission} 
The MSC framework partitions the sentence into $q$ independently encoded segments. A failure in one segment leaves the remaining $q-1$ segments intact, providing the language model with reliable context for reconstruction. Formally, $\mathbf{s}$ is divided into $q$ segments of equal character length $l_{\mathrm{MSC}} = \ell/q$. After ASCII conversion, this yields $q$ bitstreams $\{\mathbf{b}_1, \ldots, \mathbf{b}_q\}$, each of length $k = k'/q$. Each segment $\mathbf{b}_i$ is independently encoded by a $\mathcal{C}_{\mathrm{MSC}}(n, k)$ code with generator matrix $\mathbf{G}_{\mathrm{MSC}}$, producing codewords $\{\mathbf{c}_1, \ldots, \mathbf{c}_q\}$ each of length $n$. The $q$ codewords are transmitted over an AWGN channel via BPSK modulation, as described in Section~\ref{section2a}.

At the receiver, each segment $\mathbf{y}_i$ is independently decoded using OSD. We define $\alpha_{i,j} = |y_{i,j}|$ as the reliability of the $j$-th received symbol in segment $i$, and 
denote the reliability vector of segment $i$ by 
$\bm{\alpha}_i = [\alpha_{i,1}, \ldots, \alpha_{i,n}]$. As described in Section \ref{section2b}, the received segment $\mathbf{y}_i$ undergoes the permutations $\tilde{\mathbf{y}}_i = \pi_2(\pi_1(\mathbf{y}_i))$, and then all TEPs are evaluated to identify the optimal codeword. We denote the optimal decoding output of $\mathbf{y}_i$ by $\hat{\mathbf{c}}_i$, whose information bits are $\hat{\mathbf{b}}_i$. The estimates $\hat{\mathbf{b}} = [\hat{\mathbf{b}}_1, \ldots, \hat{\mathbf{b}}_q]$ are mapped by ASCII source decoding to the sentence estimate $\hat{\mathbf{s}} = \{\hat{s}_1, \ldots, \hat{s}_q\}$, with each $\hat{s}_i$ of length $l_{\mathrm{MSC}}$. The SEC module then takes $\hat{\mathbf{s}}$ as input, to reconstruct the corrupted segments and recover the full sentence $\mathbf{s}$. The MSC framework is illustrated in Fig.~\ref{fig1}.

\subsubsection{Long Code (LC) Transmission Baseline}
For comparison, we consider a conventional scheme where the complete bitstream $\mathbf{b}'$ is encoded as a single $\mathcal{C}_{\mathrm{LC}}(n', k')$ LDPC codeword and decoded via BP. Both MSC and LC schemes operate at the same code rate $R = k/n = k'/n'$, ensuring fair performance comparison. 

Despite equal code rates, the two schemes exhibit fundamentally different error characteristics. Under LC, redundancy is distributed globally; when BP decoding fails, the entire sentence is corrupted and no contextual information survives for semantic recovery. 
Under MSC, failures are localized to individual segments. Correctly decoded segments provide partial context essential for reconstruction.

\begin{remark}
This work adopts ASCII encoding as the source coding scheme. However, any fixed-rate source coding scheme that produces a binary bitstream of known length $k'$ can be employed directly in the MSC framework. For variable-rate source coding, such as Huffman or arithmetic coding, error propagation across segment boundaries should be considered, since a single corrupted bit can invalidate the source decoding of subsequent symbols. In such cases, segment boundaries should be aligned with source coding units to preserve the localized failure property that enables semantic recovery. 
\end{remark}

\begin{figure*}[t]
\centering
\includegraphics[width=0.95\textwidth]{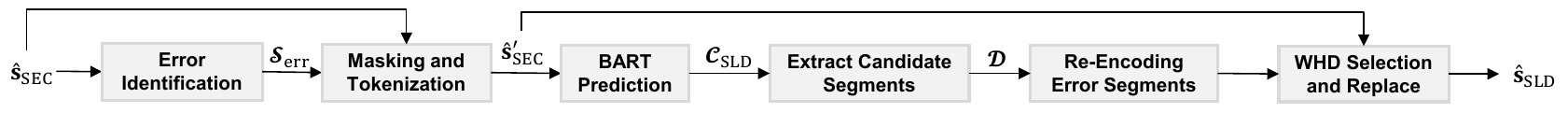}
\caption{SLD processing flow. BART generates multiple candidate segments for error segments. Candidates are re-encoded and ranked by WHD. The minimum-distance candidate is selected for each segment to construct the final output $\hat{\mathbf{s}}_{\mathrm{SLD}}$}
\vspace{-0.8em}
\label{fig:SLD_model}
\end{figure*}

\vspace{-1em}
\subsection{Semantic Error Correction (SEC)}
\label{section3b}
The sentence estimate $\hat{\mathbf{s}}$ is fed into the SEC module. SEC operates as a denoising step at the sentence level through the BART model.

\subsubsection{SEC Processing}
The input $\hat{\mathbf{s}}$ is first tokenized into a sequence of subword units using the pretrained BPE function $f_{\mathrm{BPE}}(\cdot)$ introduced in Section~\ref{section2c}: 
\begin{equation}
    \hat{\mathbf{t}} = f_{\mathrm{BPE}}(\hat{\mathbf{s}}) = [\hat{t}_1, \hat{t}_2, \ldots, \hat{t}_z],
    \blue{\label{eq:tokenize}}
\end{equation}
where $z$ is the token sequence length, which varies with sentence content. Each token $\hat{t}_i$ is a high-dimensional vector representing a subword unit from the sentence, which may be a complete word, a word fragment, or even a single character, depending on the tokenization process. When channel decoding errors introduce garbled or out-of-vocabulary characters, BPE recursively decomposes them into individual characters, ensuring the tokenizer remains well-defined under arbitrary corruption.
The token sequence $\hat{\mathbf{t}}$ is passed to the BART correction model:
\begin{equation}
    \hat{\mathbf{s}}_{\mathrm{SEC}} = f_{\mathrm{SEC}} (\hat{\mathbf{t}};\, \boldsymbol{\theta}_{\mathrm{SEC}}),
    \label{eq:sec}
\end{equation}
where $\hat{\mathbf{s}}_{\mathrm{SEC}}=\{\hat{\mathbf{s}}_{\mathrm{SEC},1},\hat{\mathbf{s}}_{\mathrm{SEC},2}\ldots,\hat{\mathbf{s}}_{\mathrm{SEC},q}\}$ is the corrected sentence and $\boldsymbol{\theta}_{\mathrm{SEC}}$ is the fine-tuned model parameters. BART's bidirectional encoder captures dependencies across all segments simultaneously, allowing correctly decoded segments to inform the reconstruction of corrupted ones.

Table~\ref{tab1} illustrates the SEC pipeline for error scenarios using $(128, 64)$ codes at $\mathrm{SNR} = 2\,\mathrm{dB}$. The full sentence is transmitted using $q=8$. Even when a segment is severely corrupted  (e.g., ``\texttt{c/?ueK(y}''), SEC recovers a semantically plausible substitution from the surrounding context. Note that the recovered segment can still be different to the original. The confidence scoring in Section~\ref{section4} can determine whether the correction is reliable.

The training process of SEC will be detailed in Section~\ref{section5train}.

\vspace{-0.8em}
\section{Proposed Semantic List Decoding Scheme}
\vspace{-0.3em}
\label{section4}

SEC relies solely on context reasoning, so its errors remain grammatically and semantically plausible and cannot be detected from text alone.
For example, as shown in Table~\ref{tab1}, if ``A lady i'' is corrupted in ``A lady in a blue shirt~\ldots'' , SEC may infer ``A girl i'' based on context.

To address this, SLD introduces two stages: error identification and list-based correction, as illustrated in Fig.~\ref{fig:SLD_model}. The error identification step detects segments where the SEC reconstruction is unreliable by comparing the re-encoded SEC output against the received signal.
For each identified unreliable segment, the list-based correction step generates multiple candidate reconstructions and selects the most reliable one using weighted Hamming distance against the received signal.

\vspace{-22pt}
\subsection{Error identification}
\label{section4a}
SEC operates at the text level, whereas the proposed error identification requires comparison against the received signal in the codeword domain. We therefore re-encode each SEC output segment back into a codeword and measure its consistency with the channel observation.

For the $i$-th segment after SEC processing, $\hat{\mathbf{s}}_{\mathrm{SEC},i}$ is first converted to its $k$-bit ASCII representation $\hat{\mathbf{b}}_{\mathrm{SEC},i}$, and then re-encoded into a length-$n$ codeword $\hat{\mathbf{c}}_{\mathrm{SEC},i} = \hat{\mathbf{b}}_{\mathrm{SEC},i} \mathbf{G}_{\mathrm{MSC}}$. Applying the OSD permutations from Section~II-B yields the permuted codeword  $\tilde{\mathbf{c}}_{\mathrm{SEC},i} = \pi_2(\pi_1(\hat{\mathbf{c}}_{\mathrm{SEC},i}))$. For
notational simplicity, the segment index $i$ is omitted in the
following, i.e., $\tilde{\mathbf{c}}_{\mathrm{SEC}} = \tilde{\mathbf{c}}_{\mathrm{SEC},i}$. The re-encoded codeword $\tilde{\mathbf{c}}_{\mathrm{SEC}}$ implicitly defines a TEP $$\mathbf{e}_{\mathrm{SEC}} = \tilde{\mathbf{r}}_B \oplus \tilde{\mathbf{c}}_{\mathrm{SEC},B}$$ on the first $k$ systematic bits. Unlike conventional OSD, which enumerates TEPs of Hamming weight up to $m$, the TEP $\mathbf{e}_{\mathrm{SEC}}$ can have arbitrary weight and originates from semantic inference rather than systematic enumeration. 

To assess whether this semantically-derived TEP is reliable, we compute its \emph{success probability} $P_{\mathrm{SEC},i}$, defined as the \emph{a~posteriori} probability that $\mathbf{e}_{\mathrm{SEC}}$ correctly identifies the true channel error pattern:
\begin{equation}
P_{\mathrm{SEC},i} \triangleq 
\Pr\left(\tilde{\mathbf{e}}_B = \mathbf{e}_{\mathrm{SEC}} \mid 
\mathbf{d}_{\mathrm{SEC}}\right),
\end{equation}
where $\mathbf{d}_{\mathrm{SEC}} = \tilde{\mathbf{c}}_{\mathrm{SEC}} \oplus \tilde{\mathbf{r}}$ is the difference pattern between the re-encoded codeword and the hard-decision received vector, and $\Pr(\tilde{\mathbf{e}}_B = \mathbf{e}_{\mathrm{SEC}})$ is the \emph{a~priori} probability of $\tilde{\mathbf{e}}_B = \mathbf{e}_{\mathrm{SEC}}$.

Let $\mathbf{D}_{\mathrm{SEC}}$ denote the random variable corresponding to $\mathbf{d}_{\mathrm{SEC}}$. For notational brevity, we write $\Pr(\mathbf{d}_{\mathrm{SEC}})$ for $\Pr(\mathbf{D}_{\mathrm{SEC}} = \mathbf{d}_{\mathrm{SEC}})$ in the following. Applying the probability analysis framework in \cite{pbosd}, $P_{\mathrm{SEC},i}$ can be expressed via Bayes' theorem as
\begin{equation}
P_{\mathrm{SEC},i} = \frac{\Pr(\mathbf{d}_{\mathrm{SEC}} | \tilde{\mathbf{e}}_B = \mathbf{e}_{\mathrm{SEC}}) \Pr(\tilde{\mathbf{e}}_B = \mathbf{e}_{\mathrm{SEC}})}{\Pr(\mathbf{d}_{\mathrm{SEC}})}.
\end{equation}
Following~\cite{pbosd}, $P_{\mathrm{SEC},i}$ can be approximated as
\begin{equation}
P_{\mathrm{SEC},i} \approx \Bigg(1 + \frac{(1-P(\mathbf{e}_{\mathrm{SEC}}))2^{k-n}} 
{P(\mathbf{e}_{\mathrm{SEC}}) \displaystyle\prod_{\substack{k \leq j \leq n \\ d_{\mathrm{SEC},j} \neq 0}} \!\!P(j) \displaystyle\prod_{\substack{k \leq j \leq n \\ d_{\mathrm{SEC},j} = 0}} \!\!(1-P(j))}\Bigg)^{\!-1}\!\!,
\label{eq:psec_approx}
\end{equation}
where $P(j) \approx (1+\exp(2\alpha_j))^{-1}$ is the bit error probability at position~$j$ given reliability~$\alpha_j$, and $P(\mathbf{e}_{\mathrm{SEC}})$ is the prior probability of the error pattern $\mathbf{e}_{\mathrm{SEC}}$ computed from the individual bit error probabilities as detailed in~\cite[Eq. (5)]{pbosd}. Segments $\hat{\mathbf{s}}_{\mathrm{SEC},i}$ with $P_{\mathrm{SEC},i} < T_{\mathrm{SEC}}$ form the error set, i.e.,
\[
\mathcal{S}_{\mathrm{err}} = \{i  : P_{\mathrm{SEC},i} < T_{\mathrm{SEC}}\}
\]
where $T_{\mathrm{SEC}} \in (0,1)$ is a confidence threshold. If all segments meet the threshold ($\mathcal{S}_{\mathrm{err}} = \emptyset$), SEC outputs are used directly without further processing.

\vspace{-1em}
\subsection{Semantic List Decoding}
\vspace{-0.3em}
SLD refines SEC outputs when error identification detects unreliable segments ($\mathcal{S}_{\mathrm{err}} \neq \emptyset$). To process each $\hat{\mathbf{s}}_{\mathrm{SEC},i} \in \mathcal{S}_{\mathrm{err}}$, we assume all other segments $\hat{\mathbf{s}}_{\mathrm{SEC},j}$ $(j\neq i)$, are correct. Then, we replace $\hat{\mathbf{s}}_{\mathrm{SEC},i}$ with the special symbol $\langle\text{mask}\rangle$ from the BPE vocabulary, yielding the masked sentence
\begin{equation}
\hat{\mathbf{s}}_{\mathrm{SEC}}' = \{\hat{\mathbf{s}}_{\mathrm{SEC},1}, \ldots, \langle\text{mask}\rangle_i, \ldots, \hat{\mathbf{s}}_{\mathrm{SEC},q}\}.
\label{eq:mask}
\end{equation}

\subsubsection{SLD Processing}
$\hat{\mathbf{s}}'_{\mathrm{SEC}}$ is tokenized to obtain\gray{:}
\begin{equation}
\hat{\mathbf{t}}' = f_{\text{BPE}}(\hat{\mathbf{s}}_{\mathrm{SEC}}') = [\hat{t}'_1, \hat{t}'_2, \ldots, \hat{t}'_{z'}],
\end{equation}
where $z'$ is the length of the masked token sequence. The token sequence $\hat{\mathbf{t}}'$ is passed to the BART model to generate $V$ candidate sentences via
\begin{equation}
\mathcal{C}_{\mathrm{SLD}} = f_{\mathrm{SLD}}(\hat{\mathbf{t}}';\bm{\theta}_{\mathrm{SLD}},V), 
\label{eq:GB}
\end{equation} 
where $\mathcal{C}_{\mathrm{SLD}} = \{\hat{\mathbf{s}}^1_{\mathrm{SLD}}, \ldots, \hat{\mathbf{s}}^V_{\mathrm{SLD}}\}$ denotes the set of $V$ candidate reconstructions and $\bm{\theta}_{\mathrm{SLD}}$ denotes the fine-tuned model parameters. Unlike SEC, which produces a single output, SLD retains the top-$V$ high-probability sequences from BART's autoregressive decoding to form a diverse candidate set. The specific strategy used to generate $\mathcal{C}_{\mathrm{SLD}}$ is detailed in Section~VII.

Generating multiple candidates, however, raises an alignment problem. BART produces each candidate one token at a time. Different candidates may use different numbers of tokens, so the reconstructed segment can appear at different character positions across $\mathcal{C}_{\mathrm{SLD}}$. Fixed-length slicing as in MSC is therefore no longer valid; we need to locate each masked segment within each candidate.

\subsubsection{Locating the Reconstructed Content}
For each masked segment $i \in \mathcal{S}_{\mathrm{err}}$, we extract its candidate segments from the $V$ candidate sentences as
\begin{equation}
\mathcal{D}_i = f_{\text{extract}}(\mathcal{C}_{\mathrm{SLD}}, \hat{\mathbf{s}}_{\mathrm{SEC}}', i),
\label{eq:extract}
\end{equation}
where each element of $\mathcal{D}_i$ has length~$l_{\mathrm{MSC}}$.

The extraction function $f_{\mathrm{extract}}(\cdot,\cdot,\cdot)$ locates the reconstructed content for each masked segment $i \in \mathcal{S}_{\mathrm{err}}$ in each candidate $\hat{\mathbf{s}}_{\mathrm{SLD}}^v \in \mathcal{C}_{\mathrm{SLD}}$. Since correctly decoded segments are preserved across all candidates, they serve as anchors that localize the reconstructed content despite the length variations. The extraction proceeds as follows.

\begin{figure}
\centering
\includegraphics[width=0.49 \textwidth]{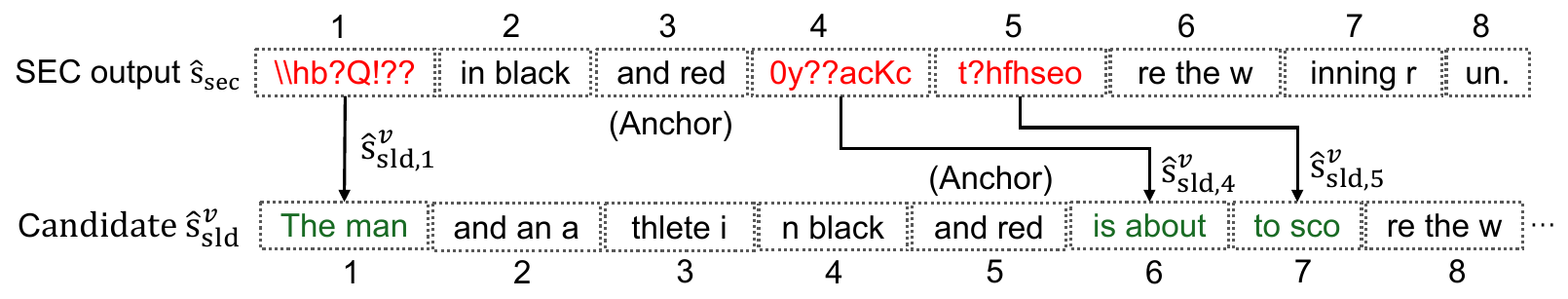}
\caption{Illustration of reconstructed contents extraction with $\mathcal{S}_{\mathrm{err}} = \{1, 4, 5\}$. Segment 1 is extracted as a leading segment. Segments 4 and 5 are consecutive masked segments.}
\label{fig:sld_candidate}
\vspace{-15pt}
\end{figure}

\begin{itemize}
    \item \textbf{Leading segment ($i=1$):} No preceding anchor exists. The first $l_{\mathrm{MSC}}$ characters are directly extracted from each candidate $\hat{\mathbf{s}}^{v}_{\mathrm{SLD}} \in \mathcal{C}_{\mathrm{SLD}}$.
    
    \item \textbf{Non-leading segment ($i > 1$):} If $i{-}1 \notin \mathcal{S}_{\mathrm{err}}$, the preceding segment $\hat{\mathbf{s}}_{\mathrm{SEC},i-1}$ serves as the anchor. The subsequent $l_{\mathrm{MSC}}$ characters are extracted.
    
    \item \textbf{Consecutive masked segments ($i, i{+}1, \ldots, i{+}r$ for $r \geq 1$):}  The nearest reliable segment preceding the group serves as the anchor, after which $r{+}1$ consecutive blocks of length $l_{\mathrm{MSC}}$ are extracted.
\end{itemize}
Figure~\ref{fig:sld_candidate} illustrates this procedure. Applying it to all $V$ candidates yields the candidate segment set $\mathcal{D}_i = \{\hat{\mathbf{s}}^1_{\mathrm{SLD},i}, \ldots, \hat{\mathbf{s}}^v_{\mathrm{SLD},i}, \ldots, \hat{\mathbf{s}}^V_{\mathrm{SLD},i}\}$.

\vspace{-0.5em}
\subsection{Weighted Hamming Distance Selection}

Since all candidates in $\mathcal{D}_i$ are semantically plausible, we perform bit-level selection by evaluating each candidate's distance against the received signal. Each candidate segment $\hat{\mathbf{s}}^v_{\mathrm{SLD},i}$ is first converted to its $k$-bit ASCII representation $\hat{\mathbf{b}}^v_{\mathrm{SLD},i}$, then re-encoded using the generator matrix $\mathbf{G}_{\mathrm{MSC}}$ to obtain the codeword $\hat{\mathbf{c}}^v_{\mathrm{SLD},i} = \hat{\mathbf{b}}^v_{\mathrm{SLD},i}\mathbf{G}_{\mathrm{MSC}}$. The weighted Hamming distance $d^{(W)}(\hat{\mathbf{c}}^{v}_{\mathrm{SLD},i}, \mathbf{y}_i)$ between $\hat{\mathbf{c}}^{v}_{\mathrm{SLD},i}$ and $\mathbf{y}_i$ is then computed following~\eqref{eq:whd_osd}.

Re-encoding maps each semantically generated candidate back to the codeword domain, enabling direct comparison with the channel observation. When the correct segment is contained in $\mathcal{D}_i$, its re-encoded codeword typically achieves the lowest WHD, as it is closest to the transmitted codeword. For each erroneous segment $i \in \mathcal{S}_{\mathrm{err}}$, we select the candidate minimizing the WHD:
\begin{equation}
\hat{\mathbf{s}}^{\mathrm{opt}}_{\mathrm{SLD},i} = \arg\min_{\hat{\mathbf{s}}^{v}_{\mathrm{SLD},i} \in \mathcal{D}_i} d^{(W)}(\hat{\mathbf{c}}^{v}_{\mathrm{SLD},i}, \mathbf{y}_i).
\label{eq:vopt}
\end{equation}

\begin{figure}
\centering
\includegraphics[width=0.45\textwidth]{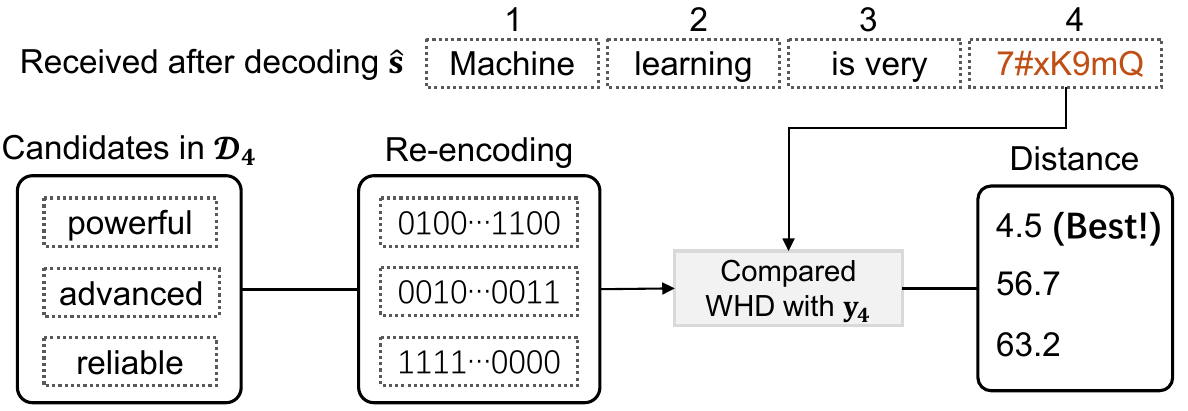}
    \caption{Example of WHD selection in SLD: candidates in $\mathcal{D}_4$ are re-encoded and ranked by their WHD against $\mathbf{y}_4$; "powerful" achieves the minimum distance and is selected.}
\vspace{-0.8em}
\label{fig:SLD_highlight}
\end{figure}

\begin{algorithm}[t]
\small
\caption{MSC Receiver with SEC and SLD}
\label{alg:msc_sld}
\begin{algorithmic}[1]
\REQUIRE Received signal $\mathbf{y} = [\mathbf{y}_1, \ldots, \mathbf{y}_q]$, generator matrix $\mathbf{G}_{\mathrm{MSC}}$, confidence threshold $T_{\mathrm{SEC}}$, candidate list size $V$
\ENSURE Decoded sentence $\hat{\mathbf{s}}$
\vspace{2pt}
\FOR{$i = 1, \ldots, q$ \textbf{in parallel}} 
    \STATE $\hat{\mathbf{b}}_i \leftarrow \mathrm{OSD}(\mathbf{y}_i)$  \algcomment{Stage 1: Parallel Channel Decoding}
    \STATE $\hat{\mathrm{s}}_i \leftarrow f_{\mathrm{ASCII}}^{-1}(\hat{\mathbf{b}}_i)$
\ENDFOR
\vspace{2pt}
\STATE $\hat{\mathbf{s}}_{\mathrm{SEC}} \leftarrow f_{\mathrm{SEC}}(f_{\mathrm{BPE}}(\hat{\mathbf{s}}); \boldsymbol{\theta}_{\mathrm{SEC}})$ \algcomment{Stage 2: Semantic Error Correction}
\FOR{$i = 1, \ldots, q$}
    \STATE $\hat{\mathbf{c}}_{\mathrm{SEC},i} \leftarrow f_{\mathrm{ASCII}}(\hat{\mathrm{s}}_{\mathrm{SEC},i}) \, \mathbf{G}_{\mathrm{MSC}}$ \algcomment{Re-encoding}
    \STATE Compute $P_{\mathrm{SEC},i}$ via~\eqref{eq:psec_approx}
\ENDFOR
\STATE $\mathcal{S}_{\mathrm{err}} \leftarrow \{i : P_{\mathrm{SEC},i} < T_{\mathrm{SEC}}\}$ \algcomment{Error Identification}
\IF{$\mathcal{S}_{\mathrm{err}} = \emptyset$}
    \RETURN $\hat{\mathbf{s}}_{\mathrm{SEC}}$
\ENDIF
\vspace{2pt}
\STATE Form $\hat{\mathbf{s}}'_{\mathrm{SEC}}$ for all $i \in \mathcal{S}_{\mathrm{err}}$ via \eqref{eq:mask}  \algcomment{Stage 3: List Decoding}
\STATE Generate V candidates $\mathcal{C}_{\mathrm{SLD}}$ via \eqref{eq:GB} 
\FOR{each $i \in \mathcal{S}_{\mathrm{err}}$}
    \STATE Extract candidate set $\mathcal{D}_i$ via~\eqref{eq:extract}
    \STATE Re-encode each $\hat{\mathrm{s}}^v_{\mathrm{SLD},i} \in \mathcal{D}_i$ into $\hat{\mathbf{c}}^v_{\mathrm{SLD},i}$
    \STATE Evaluate $d^{(W)}(\hat{\mathbf{c}}^v_{\mathrm{SLD},i}, \mathbf{y}_i)$ via~\eqref{eq:whd_osd}
    \STATE Select $\hat{\mathrm{s}}^{\mathrm{opt}}_{\mathrm{SLD},i}$ via~\eqref{eq:vopt}
\ENDFOR
\RETURN $\hat{\mathbf{s}}_{\mathrm{SLD}}$ via~\eqref{eq::SLD final}
\end{algorithmic}
\end{algorithm}

The final SLD output is constructed by replacing each erroneous segment with its optimal candidate while preserving SEC outputs at other segments:
\begin{equation} \label{eq::SLD final}
\hat{\mathbf{s}}_{\mathrm{SLD}} = \left\{
\begin{array}{ll}
\hat{\mathbf{s}}^{\mathrm{opt}}_{\mathrm{SLD},i} & \text{if } i \in \mathcal{S}_{\text{err}}, \\
\hat{\mathbf{s}}_{\mathrm{SEC},i} & \text{otherwise}.
\end{array}
\right. 
\end{equation}
Only segments flagged as unreliable are refined by SLD. Figure~\ref{fig:SLD_highlight} illustrates this selection process for an example segment.

\vspace{-1em}

\subsection{Algorithm}
Algorithm~\ref{alg:msc_sld} summarizes the complete MSC receiver. It proceeds in three stages: parallel OSD decoding of the $q$ segments, SEC with error identification, and SLD refinement of the identified erroneous segments.

\vspace{-0.5em}
\section{Semantic Retransmission Scheme}
\vspace{-0.3em}
\label{section5}
Even with SLD, some segments may remain unreliable when both semantic context and channel information are insufficient. This can be further resolved by retransmission. Conventional HARQ detects errors via CRC and retransmits the entire message or additional parity-check bits upon failure. We propose SHARQ, which retransmits only the segments identified as unreliable.
\vspace{-1em}
\subsection{Error Identification}
\vspace{-0.3em}
Even after SLD selection, $\hat{\mathbf{s}}^{\mathrm{opt}}_{\mathrm{SLD},i} \neq \mathbf{s}_i$ remains possible when (i) the true segment $\mathbf{s}_i$ is not generated by the language model, so $\mathbf{s}_i \notin \mathcal{D}_i$, or (ii) the channel reliability $\boldsymbol{\alpha}_i$ is too low for WHD to discriminate among candidates. We therefore compute $P_{\mathrm{SLD},i}$, defined analogously to~\eqref{eq:psec_approx} but evaluated at the re-encoded optimal candidate $\hat{\mathbf{c}}^{\mathrm{opt}}_{\mathrm{SLD},i} = f_{\mathrm{ASCII}}(\hat{\mathbf{s}}^{\mathrm{opt}}_{\mathrm{SLD},i})\,\mathbf{G}_{\mathrm{MSC}}$. As $P_{\mathrm{SLD},i}$ combines the language-model-induced prior (through $\mathcal{D}_i$) with the channel soft information (through $\boldsymbol{\alpha}_i$), it serves as a joint semantic–channel confidence score. Segments with confidence below a threshold $T_{\mathrm{harq}} \in (0,1)$ are marked for retransmission,
\[
\mathcal{S}_{\mathrm{harq}} = \{i : P_{\mathrm{SLD},i} < T_{\mathrm{harq}}\}.
\]

\subsection{Semantic Confidence-Guided HARQ}

When $\mathcal{S}_{\mathrm{harq}} \neq \emptyset$, the receiver requests retransmission of the segments indexed by $\mathcal{S}_{\mathrm{harq}}$. When the retransmission budget is constrained to $M < |\mathcal{S}_{\mathrm{harq}}|$ segments, the receiver requests the $M$ segments with the lowest confidence (i.e., $P_{\mathrm{SLD},i}$). 
Following incremental redundancy (IR) HARQ, the transmitter sends additional parity bits for each requested segment, received as $\mathbf{y}^*_i$.The receiver concatenates the retransmitted and original observations,
\begin{equation}
\mathbf{y}_i^r = [\mathbf{y}_i, \mathbf{y}_i^{*}], \quad i \in \mathcal{S}_{\mathrm{harq}},
\end{equation}
and decodes them via OSD and ASCII source decoding to obtain the updated segment estimates $\hat{\mathbf{s}}^r_i$.

The updated sentence estimate is assembled as
\begin{equation}
{\hat{\mathbf{s}}}^r = \begin{cases}
\hat{\mathbf{s}}_i^r & \text{if } i \in \mathcal{S}_{\mathrm{harq}} \\
\hat{\mathbf{s}}_{\mathrm{SLD},i} & \text{otherwise}.
\end{cases}
\end{equation}

The assembled $\hat{\mathbf{s}}^r$ is then re-processed through the full SEC and SLD pipeline, which produces a new error set $\mathcal{S}^r_{\mathrm{harq}}$. Retransmission rounds can continue until $\mathcal{S}^r_{\mathrm{harq}} = \emptyset$ or the maximum number of rounds is reached.

\vspace{-0.5em}
\subsection{Elimination of CRC Overhead}
Conventional HARQ appends a $k_{\mathrm{crc}}$-bit CRC to each segment's information bits for error detection, reducing the effective code rate from $R = k/n$ to $(k-k_{\mathrm{crc}})/n$. The overhead ratio $\rho_{\mathrm{crc}} = k_{\mathrm{crc}}/k$ is negligible for long codes. For example, a 16-bit CRC on a $(1024, 512)$ codeword yields $\rho_{\mathrm{crc}} = 3.1\%$. However, it becomes prohibitive for short codes: an 8-bit CRC on a $(32, 16)$ code halves the throughput ($\rho_{\mathrm{crc}} = 50\%$).

SHARQ eliminates this overhead by replacing the CRC with $P_{\mathrm{SLD},i}$, which is computed directly from the channel reliabilities $\boldsymbol{\alpha}$ and carries no transmitted redundancy. The effective rate therefore equals the nominal rate $R$, and the throughput gain over CRC-HARQ is
\begin{equation}
G = \frac{k}{k - k_{\mathrm{crc}}} - 1 = \frac{k_{\mathrm{crc}}}{k - k_{\mathrm{crc}}},
\label{eq:sharq_gain}
\end{equation}
independent of the number of segments $q$.

\vspace{-0.5em}
\section{Performance and Complexity Analysis}
\vspace{-0.3em}
\subsection{SEC Analysis}
\vspace{-0.3em}
\subsubsection{Analytical Framework}
Let $\mathcal{E} \subseteq \{1,\ldots,q\}$ denote the set of segments that are erroneous after OSD decoding, and let $q_e = |\mathcal{E}|$. For segment $i \in \mathcal{E}$, let $R_i$ denote the event that the semantic module successfully recovers segment~$i$. In general, $\mathbb{P}(R_i \mid i \in \mathcal{E})$ depends on which segments fail and on their semantic content, which makes a closed-form analysis intractable. Therefore, we adopt the following simplification.

\noindent\textit{Assumption 1:} The recovery probability of any erroneous segment depends on the error pattern $\mathcal{E}$ only through its cardinality $q_e = |\mathcal{E}|$. We accordingly define

\begin{equation}
P_{\mathrm{rec}}(q_e) \triangleq \mathbb{P}(R_i \mid i \in \mathcal{E},\, |\mathcal{E}| = q_e),
\label{P_rec}
\end{equation}
which is independent of the segment index $i$ and of the specific positions of the other erroneous segments.

Table~\ref{tab:prec} reports $P_{\mathrm{rec}}(q_e)$ measured at SNR $= 1.5$~dB for $q_e \leq 4$, with $P_{\mathrm{rec}}(0) \triangleq 1$ since no recovery is needed when all segments decode correctly. Entries for $q_e = 4$ on longer codes such as $(128,64)$ are omitted, as the event is exceedingly rare at this SNR. As seen, $P_{\mathrm{rec}}(q_e)$ decreases monotonically with $q_e$ for every code configuration, reflecting the loss of contextual segments available for reconstruction. Furthermore, at the same $q_e$, shorter codes yield higher recovery probabilities, because each failure spans fewer characters.

\begin{table}[t]
\centering
\caption{Empirical recovery probability $P_{\text{rec}}(q_e)$ for $q_e \leq 4$ at SNR $= 1.5$~dB.}
\label{tab:prec}
\begin{tabular}{cl|ccc}
\hline
& & $\mathcal{C}(64,32)$ & $\mathcal{C}(128,64)$ & $\mathcal{C}(256,128)$ \\
& $q_e$ & $q=16$ & $q=8$ & $q=4$ \\
\hline
& 0 & 1 & 1 & 1 \\
& 1 & 0.898 & 0.635 & 0.227 \\
SEC & 2 & 0.827 & 0.525 & 0.250 \\
& 3 & 0.735 & 0.111 & -- \\
& 4 & 0.608 & -- & -- \\
\hline
\end{tabular}
\vspace{-2em}
\end{table}

The statistical recovery rate
\begin{equation}
\eta = \sum_{q_e=1}^{q} P_{\mathrm{rec}}(q_e) \cdot \mathbb{P}(|\mathcal{E}|=q_e ),
\end{equation}
is a weighted average of $P_{\mathrm{rec}}(q_e)$ over all error multiplicities.

Since the $q$~segments are independently decoded and experience i.i.d.\ channel noise, the number of erroneous segments follows a binomial distribution:
\begin{equation}
\mathbb{P}(|\mathcal{E}|=q_e) = \binom{q}{q_e} (P_e^{\mathrm{OSD}})^{q_e} (1 - P_e^{\mathrm{OSD}})^{q-q_e},
\end{equation}
where $P_e^{\mathrm{OSD}}$ is the per-segment OSD block error probability. For an order-$m$ OSD applied to a $\mathcal{C}(n,k)$ code, a decoding error occurs when the number of channel errors in the $k$~most reliable positions exceeds~$m$. Let $p_{E_1^k}(j)$ denote the probability that exactly $j$ of these $k$ positions are in error, as characterized by~\cite[Lemma~1]{pbosd}. Then
\begin{equation}
P_e^{\mathrm{OSD}} \approx \sum_{j=m+1}^{k} p_{E_1^k}(j) + P_{\mathrm{ML}},
\end{equation}
where $p_{E_1^k}(j)$ is given by~\cite[Eq.~(12)]{pbosd} and depends on the SNR through the bit error probabilities of the ordered reliability positions. Here, $P_{\mathrm{ML}}$ is the maximum-likelihood error probability of $\mathcal{C}_{\mathrm{MSC}}(n, k)$, which is an inherent property of the code. Therefore, we have
\begin{equation}
\eta = \sum_{q_e=1}^{q} P_{\mathrm{rec}}(q_e) \cdot \binom{q}{q_e} (P_e^{\mathrm{OSD}})^{q_e} (1 - P_e^{\mathrm{OSD}})^{q-q_e}.
\end{equation}

To derive the sentence-level BLER, we require one additional assumption.

\noindent\textit{Assumption 2:}
Given the number of erroneous segments $q_e$ and the channel observations, the recovery outcomes of individual erroneous segments are mutually independent, each succeeding with probability $P_{\mathrm{rec}}(q_e)$.

Under this assumption, the probability that all $q_e$ erroneous segments are simultaneously recovered is $[P_{\mathrm{rec}}(q_e)]^{q_e}$. Averaging over the binomial distribution of $q_e$ yields
\begin{equation}
\mathrm{BLER} =1- \sum_{q_e=0}^{q} \binom{q}{q_e} (P_e^{\mathrm{OSD}})^{q_e}(1-P_e^{\mathrm{OSD}})^{q-q_e}\, [P_{\mathrm{rec}}(q_e)]^{q_e},
\label{eq:cal_bler}
\end{equation}
If $P_{\mathrm{rec}}(q_e)$ is approximated by $\eta$ independent of $q_e$, we have
\begin{equation}
\mathrm{BLER} \approx 1 - \big(1 - P_e^{\mathrm{OSD}}(1 - \eta)\big)^q.
\label{eq:est_bler}
\end{equation}

\vspace{-1em}
\subsection{Extension to SLD}
The above analytical framework cannot directly apply to SLD. The recovery event for SLD decomposes as
\begin{equation}
P_{\mathrm{rec}}^{\mathrm{SLD}}(q_e) = P_{\mathrm{cover}}(q_e) \cdot P_{\mathrm{select}}(q_e).
\end{equation}

Given a candidate set, $P_{\mathrm{select}}$ reduces to a minimum-distance decoding problem over $|\mathcal{D}_i|$ codewords and is amenable to channel-coding analysis. The coverage probability $P_{\mathrm{cover}}$, however, depends on the language model's distribution, the candidate diversity, and the available semantic context. These factors are determined by training data and model architecture rather than by channel or coding parameters.

\vspace{-1em}
\subsection{Information-Theoretic Interpretation}
We provide an information-theoretic interpretation of the proposed framework. Let $\mathcal{E} \subseteq \{1,\ldots,q\}$ denote the set of erroneous segments after channel decoding, and let $\mathbf{s}_{\mathcal{E}} = \{s_i\}_{i \in \mathcal{E}}$ and $\mathbf{s}_{\bar{\mathcal{E}}} = \{s_j\}_{j \notin \mathcal{E}}$ denote the erroneous and correctly decoded segments, respectively.

\subsubsection{Residual uncertainty after decoding failure} Under LC, a decoding failure corrupts the entire sentence. The receiver retains only the channel observation $\mathbf{y}$, so the residual uncertainty is
\begin{equation}
H_{\mathrm{LC}} \triangleq H(\mathbf{s} \mid \mathbf{y},  \mathcal{F}_{\mathrm{LC}}) = H(\mathbf{s} \mid \mathbf{y}).
\end{equation}
where $\mathcal{F}_{\mathrm{LC}}$ denotes the LC decoding failure event. Under MSC, the correctly decoded segments $\mathbf{s}_{\bar{\mathcal{E}}}$ are known exactly, and the residual uncertainty is
\begin{equation}
H_{\mathrm{MSC}} \triangleq H(\mathbf{s} \mid \mathbf{y}, \mathbf{s}_{\bar{\mathcal{E}}}) = H(\mathbf{s}_{\mathcal{E}} \mid \mathbf{y}, \mathbf{s}_{\bar{\mathcal{E}}}),
\end{equation}
where the equality follows because $\mathbf{s}_{\bar{\mathcal{E}}}$ is a deterministic function of $\mathbf{s}$, so conditioning on it eliminates uncertainty in the known segments.

\subsubsection{Information gain from correctly decoded segments} The reduction in uncertainty provided by MSC over LC is
\begin{align}
H_{\mathrm{LC}} - H_{\mathrm{MSC}} &= H(\mathbf{s} \mid \mathbf{y}) - H(\mathbf{s}_{\mathcal{E}} \mid \mathbf{y}, \mathbf{s}_{\bar{\mathcal{E}}}) \notag \\
&= H(\mathbf{s}_{\bar{\mathcal{E}}} \mid \mathbf{y}) + I(\mathbf{s}_{\bar{\mathcal{E}}}; \mathbf{s}_{\mathcal{E}} \mid \mathbf{y}), \label{eq:info_gain}
\end{align}
where the first term $H(\mathbf{s}_{\bar{\mathcal{E}}} \mid \mathbf{y})$ accounts for the uncertainty in $\mathbf{s}_{\bar{\mathcal{E}}}$ that is resolved by successful decoding, and the second term $I(\mathbf{s}_{\bar{\mathcal{E}}};\, \mathbf{s}_{\mathcal{E}} \mid \mathbf{y})$ is the mutual information between correctly decoded and erroneous segments given the channel observation. Both terms are non-negative, thus
\begin{equation}
H_{\mathrm{MSC}} \leq H_{\mathrm{LC}}.
\label{eq:entropy_ordering}
\end{equation}
The inequality is strict whenever $\mathcal{E} \neq \{1,\ldots,q\}$, i.e., at least one segment is correctly decoded. The mutual information term $I(\mathbf{s}_{\bar{\mathcal{E}}};\, \mathbf{s}_{\mathcal{E}} \mid \mathbf{y})$ captures the \emph{semantic side information} that the language model exploits; rigorous characterization of this quantity for natural language remains open.

\subsubsection{ Connection to recovery probability} Fano's inequality links the conditional entropy and the recovery error probability. Let $P_{e,i} = \Pr(\hat{s}_i \neq s_i \mid i \in \mathcal{E})$ denote the probability that the semantic module fails to recover segment $i$. Then
\begin{equation}
H(s_i \mid \mathbf{y}, \mathbf{s}_{\bar{\mathcal{E}}}) \leq 1 + P_{e,i} \log_2(|\mathcal{S}_i| - 1), \label{eq:fano}
\end{equation}
where $|\mathcal{S}_i| =  2^k$ is the cardinality of the segment alphabet, since each segment carries $k$ information bits. Equivalently,
\begin{equation}
P_{e,i} \geq \frac{H(s_i \mid \mathbf{y}, \mathbf{s}_{\bar{\mathcal{E}}}) - 1}{\log_2(|\mathcal{S}_i| - 1)} \approx \frac{H(s_i \mid \mathbf{y}, \mathbf{s}_{\bar{\mathcal{E}}}) - 1}{k} \label{eq:fano_lb}
\end{equation}
where we used $\log_2(2^k - 1) \approx k$ for $k \geq 1$. The bound shows that the recovery error probability is at least proportional to the ratio of the residual conditional entropy to the segment length. As $q_e$ increases, fewer segments are available in $\mathbf{s}_{\bar{\mathcal{E}}}$, so $H(s_i \mid \mathbf{y}, \mathbf{s}_{\bar{\mathcal{E}}})$ is non-decreasing in $q_e$. The lower bound on $P_{e,i}$ therefore tightens with $q_e$, consistent with the empirical decrease of $P_{\mathrm{rec}}$ in Table~\ref{tab:prec}.

\subsubsection{Segmentation tradeoff} 
The Fano bound in~\eqref{eq:fano_lb} suggests a tradeoff in choosing the segmentation factor $q$ for a fixed total information length $k' = qk$. Two competing effects act on the Fano ratio $(H(s_i \mid \mathbf{y}, \mathbf{s}_{\bar{\mathcal{E}}}) - 1)/k$ as $q$ varies.

We decompose the conditional entropy as
\begin{equation} H(s_i \mid \mathbf{y}, \mathbf{s}_{\bar{\mathcal{E}}}) = H(s_i \mid \mathbf{y}) - I(s_i;\, \mathbf{s}_{\bar{\mathcal{E}}} \mid \mathbf{y}). \label{eq:decomp_mi} 
\end{equation}

The second term $I(s_i;\, \mathbf{s}_{\bar{\mathcal{E}}} \mid \mathbf{y})$ captures the \emph{semantic gain} from context. For a fixed context fraction $(q-q_e)/q$, increasing $q$ shrinks each segment to fewer characters while the relative amount of context grows in proportion. Since natural language exhibits strong inter-segment dependencies, this favors a larger mutual information per erroneous segment, which lowers the conditional entropy and the Fano ratio.

The opposing effect is the finite-blocklength penalty. As $q$ increases, each segment uses a shorter code $\mathcal{C}(k'/(qR),\, k'/q)$ operating deeper in the finite-blocklength regime. By the normal approximation~\eqref{equ::PPV}, the per-segment OSD error probability $P_e^{\mathrm{OSD}}(k'/q)$ grows with $q$, so $\mathbb{E}[q_e] = q \cdot P_e^{\mathrm{OSD}}(k'/q)$ rises and the context set $|\bar{\mathcal{E}}| = q - q_e$ shrinks. This reduces the mutual information that the semantic gain relies on.

The optimal $q^*$ minimizing the BLER in~\eqref{eq:cal_bler} therefore balances these two effects. Its value depends on both the operating SNR and the entropy structure of the source language, which is examined empirically in Section~\ref{section4d}.

\vspace{-0.5em}
\subsection{Complexity of Semantic Module}
\vspace{-0.2em}
Let $z$ denote the token sequence length after BPE tokenization. Since BART adopts the standard Transformer architecture~\cite{NIPS2017_3f5ee243}, its computational 
complexity follows directly from the per-layer analysis therein. We denote the number of encoder and decoder layers by $L_e$ and $L_d$, the hidden dimension by $d$, and the vocabulary size by $l_{\mathrm{voc}}$.

\subsubsection{BART Encoder Complexity}
Following the Transformer architecture~\cite{NIPS2017_3f5ee243}, each encoder layer has per-layer complexity $\mathcal{O}(z^2 d)$ for self-attention and $\mathcal{O}(zd^2)$ for the linear projections and feed-forward network. The total encoder 
complexity is
\begin{equation}
\mathcal{O}_{\text{enc}} \approx L_e (z^2 d + zd^2).
\end{equation}

\subsubsection{BART Decoder Complexity}
In addition to the self-attention and feed-forward components shared with the encoder, each decoder layer includes encoder-decoder cross-attention with $\mathcal{O}(z^2 d)$ cost and a vocabulary projection with $\mathcal{O}(z l_{\mathrm{voc}} d)$ cost. The total decoder complexity is
\begin{equation}
\mathcal{O}_{\text{dec}} \approx L_d (z^2 d + z d^2 + z l_{\mathrm{voc}} d).
\end{equation}

\subsubsection{SEC and SLD Complexity}
SEC generates a single output sequence, so the encoder and decoder each execute once. The overall complexity is
\begin{equation}
\mathcal{O}_{\mathrm{SEC}} = \mathcal{O}_{\text{enc}} 
+ \mathcal{O}_{\text{dec}}.
\end{equation}

SLD maintains a candidate set of size $V$ throughout the autoregressive token generation, enlarging the decoder cost by a factor of $V$. Since all candidates share the same encoder output, the overall complexity is
\begin{equation}
\mathcal{O}_{\mathrm{SLD}} = \mathcal{O}_{\text{enc}} 
+ V \cdot \mathcal{O}_{\text{dec}}.
\end{equation}

\subsubsection{Comparison to Channel Decoding}
The semantic module and the channel decoder exhibit fundamentally different complexity structures. Order-$m$ OSD has complexity $\mathcal{O}(k^{m+2})$ per segment~\cite{yue2025guesswork}, which grows rapidly with the decoding order $m = \lceil d_{\min}/4 \rceil$ and the information length $k$. In contrast, the semantic module complexity is polynomial in the token sequence length $z$ and the model dimension $d$, independent of the code parameters $(n,k)$.

Furthermore, OSD is applied independently to each of the $q$ segments and can be fully parallelized, whereas the semantic module processes the entire sentence in a single pass to exploit cross-segment context. Consequently, the semantic overhead is a fixed per-sentence cost that does not scale with $q$.

\vspace{-0.5em}
\section{Experimental Results and Discussion}
\label{section7}
\subsection{Implementation Details}
\subsubsection{Dataset and Training} \label{section5train}
We use the Stanford Natural Language Inference (SNLI) corpus \cite{snli}, selecting 20,000 sentences for training and 500 for testing. Sentence character lengths $\ell$ range from 57 to 64; zero-padding is applied to ensure fixed 512-bit inputs (64 bytes) after ASCII encoding.

For SEC training, we construct a dataset $\mathcal{D}_{\mathrm{SEC}} = \{(\mathbf{s}^{(j)}, \hat{\mathbf{s}}^{(j)})\}_{j=1}^{N}$, where each pair consists of an original sentence $\mathbf{\mathbf{s}}^{(j)}$ and its
corrupted estimate $\hat{\mathbf{\mathbf{s}}}^{(j)}$ obtained after channel encoding, AWGN transmission, and OSD decoding. Each sentence is transmitted 10 times at various SNRs with independent noise realizations, yielding $N = 200{,}000$ training pairs. For SLD training, we construct $\mathcal{D}_{\mathrm{SLD}} = \{(\mathbf{s}^{(j)}, \mathbf{s}_{\text{mask}}^{(j)})\}_{j=1}^{N}$ from the SEC training data by replacing erroneous segments with $\langle\text{mask}\rangle$ tokens, generating 200,000 masked sentence pairs.

Both models are fine-tuned from BART-base by minimizing the token-level cross-entropy
\begin{equation}
    \mathcal{L} = - \frac{1}{N} \sum_{j=1}^{N} \sum_{t=1}^{T_j} \log P\!\left(\mathbf{s}^{(j)}_t \,\big|\, \mathbf{s}^{(j)}_{<t},\, \tilde{\mathbf{s}}^{(j)}\right),
\end{equation}
where $\tilde{\mathbf{s}}^{(j)}$ is the input sentence of the $j$-th training pair, namely $\hat{\mathbf{s}}^{(j)}$ for SEC and $s^{(j)}_{\mathrm{mask}}$ for SLD. Then, $\mathbf{s}^{(j)}_t$ denotes the $t$-th target token of the reference sentence $\mathbf{s}^{(j)}$ of length $T_j$. The conditional probability is produced by BART's autoregressive decoder.

The AWGN channel and all channel coding/decoding pipelines are implemented in Sionna \cite{sionna}. For the $(256,128)$ code, OSD decoding is computationally prohibitive, so the corresponding BLER curve is replaced by the normal approximation bound \cite{PPV}. SEC and SLD share the same BART-base backbone \cite{bart}, fine-tuned on a single NVIDIA A10G GPU. SLD candidate generation uses diverse beam search \cite{beam_search} with $V = 20$ candidates partitioned into $G = 4$ groups and diversity strength $\lambda = 0.8$. All hyperparameters are listed in Table~\ref{table:sim_params}.

\begin{figure*}[htbp]
    \centering
    \includegraphics[width=0.8\textwidth]{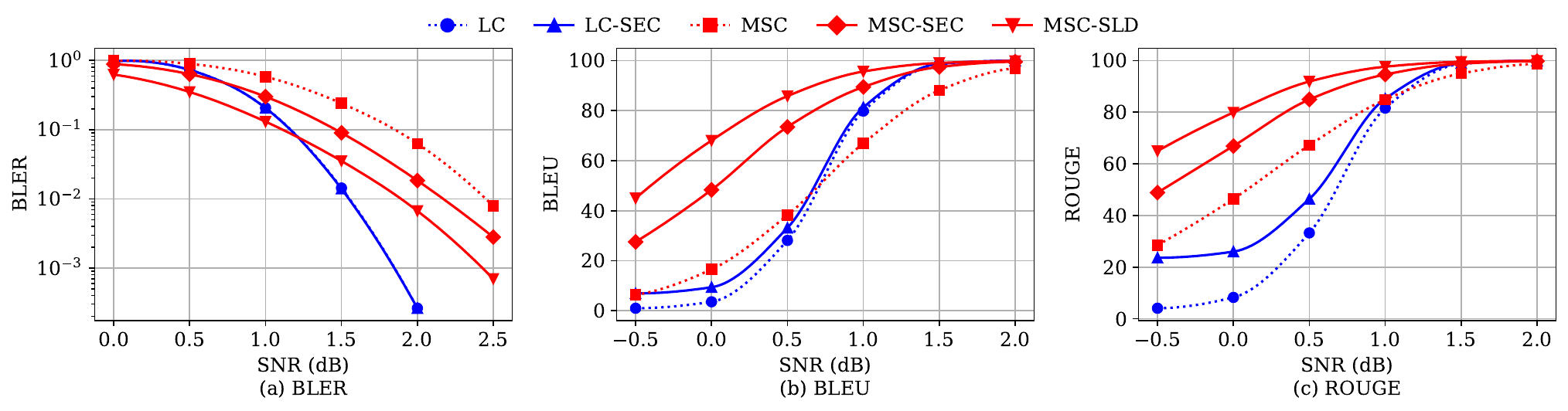}
    \vspace{-13pt}
    \caption{Performance of proposed (128,64) MSC scheme and two pipeline comapred to LC scheme in BLER, BLEU, and ROUGE.}
    \label{fig_overall}
    \vspace{-1em}
\end{figure*}

\subsubsection{Benchmarks}
We compare the following methods:
\begin{itemize}
    \item \textbf{LC:} LDPC code with $(n',k') = (1024, 512)$, decoded via BP with 80 iterations.
    \item \textbf{MSC:} We evaluate codes $\mathcal{C}(n,k)$ with $(n,k) \in \{(32,16), (64,32), (128,64), (256,128)\}$. A 512-bit sentence is segmented into $q \in \{32, 16, 8, 4\}$ parts for independent encoding. Each segment is decoded using OSD with order $m = \lfloor d_H/4 - 1 \rfloor \in \{2,2,4,8\}$.
    \item \textbf{MSC-SEC:} MSC followed by the SEC module
    \item \textbf{MSC-SLD:} MSC followed by the SEC and SLD module
\end{itemize}

For HARQ evaluation, MSC schemes use CRC-aided polar codes decoded by OSD, while LC uses CRC-aided LDPC codes decoded by BP. Both adopt a mother code of rate $R_m = 0.25$, and transmit half of the mother codeword in the initial transmission (i,e,, rate 0.5). LC selects the transmitted bits according to the 5G NR LDPC rate-matching pattern~\cite{etsi38212}, while MSC uses the most reliable positions in the polar reliability sequence~\cite{etsi38212}. The remaining bits of the mother code are delivered in the retransmission.

\begin{table}[t]
\centering
\footnotesize
\caption{Training parameters}
\renewcommand{\arraystretch}{1}
\begin{tabular}{ll|ll}
\toprule
\textbf{Parameter} & \textbf{Value} & \textbf{Parameter} & \textbf{Value} \\
\midrule
Training SNRs & $-2$ to $2$ dB & Learning rate & $3 \times 10^{-5}$ \\
Optimizer & Adam & Batch size & 128 \\
$T_{\mathrm{SEC}}$& 0.001 & $T_{\mathrm{SLD}}$& 0.1\\
\bottomrule
\end{tabular}
\label{table:sim_params}
\vspace{-15pt}
\end{table}

\subsubsection{Evaluation metrics}
Our evaluation captures both transmission reliability and semantic fidelity. For transmission reliability, we use sentence-level BLER: each sentence $\mathbf{s}$ is treated as a single block regardless of encoding method, and a block error occurs when the recovered sentence differs from the original after all processing stages.
This ensures fair comparison, as computing BLER per segment for MSC would yield misleadingly low error rates. For semantic fidelity, we adopt BLEU \cite{bleu} and ROUGE-L \cite{rouge}, which evaluate n-gram precision and longest common subsequence similarity, respectively, providing fine-grained assessments of reconstruction accuracy at the character level. In the remainder of the paper, we refer to ROUGE-L simply as ROUGE.

\vspace{-1em}

\subsection{Single Transmission Performance Comparison}
\vspace{-0.3em}
\label{section4c}
This subsection compares LC, MSC, MSC-SEC, and MSC-SLD. To isolate the effect of segmentation from that of semantic processing, we also evaluate LC augmented with the same fine-tuned SEC module, denoted LC-SEC.

\begin{figure*}[t]
\centering
\includegraphics[width=0.8\linewidth]{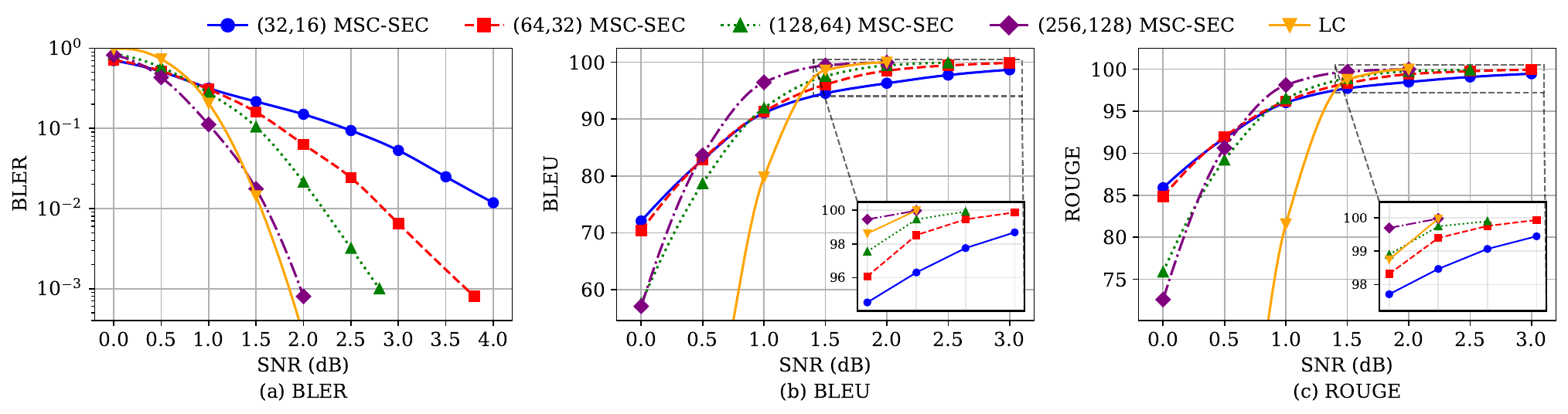}
\vspace{-13pt}
\caption{MSC-SEC performance with different code lengths in BLER, BLEU, and ROUGE.}
\label{fig:sec}
\vspace{-12pt}
\end{figure*}

\begin{figure*}[t]
\centering
\includegraphics[width=0.8\linewidth]{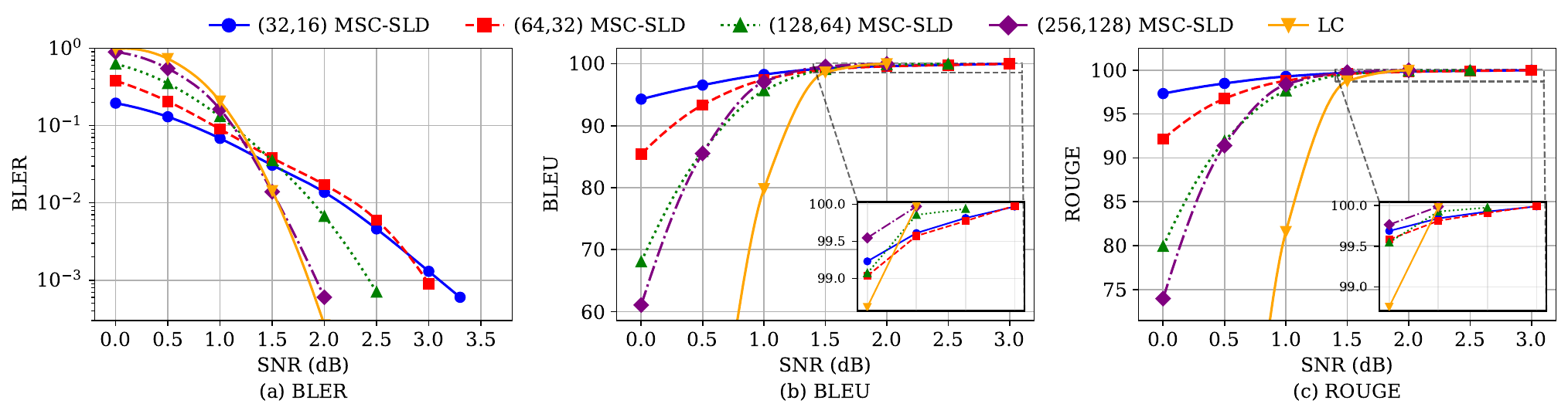}
\vspace{-13pt}
\caption{MSC-SLD performance with different code lengths in BLER, BLEU, and ROUGE.}
\label{fig:sld}
\vspace{-14pt}
\end{figure*}

\subsubsection{BLER} Figure~\ref{fig_overall}(a) compares the BLER performance for $q=8$. In the low-SNR regime, both LC and MSC exhibit high BLER. LC enters its waterfall region near $1$ dB and reaches BLER $= 10^{-4}$ at $2.1$ dB, whereas MSC reaches BLER $= 10^{-2}$ at $2.5$ dB. The gap reflects the reduced coding gain of short block codes at finite blocklength, as indicated by the normal approximation bound \cite{PPV}. Adding SEC affects the two schemes differently. For LC, LC-SEC yields negligible BLER improvement, since a decoding failure of a single long codeword typically corrupts the entire sentence and leaves no reliable context for semantic reconstruction. MSC-SEC provides a $0.4$ dB gain over MSC across all SNRs, because localized segment failures leave the surrounding segments correct for reconstruction. MSC-SLD further improves BLER, with approximate $0.8$ dB over MSC and $0.4$ dB over MSC-SEC, owing to bit-level WHD verification of multiple candidates.

\subsubsection{Semantic Fidelity}
BLEU and ROUGE scores are reported in Fig.~\ref{fig_overall}(b) and (c). At $\text{SNR} = -0.5~\text{dB}$, MSC-SEC raises BLEU from $6$ to $28$ and ROUGE from $5$ to $48$ relative to MSC; MSC-SLD raises them further to $45$ and $65$, respectively. However, the SEC model only improves the semantic scores of LC slightly. The segmentation strategy of MSC-series methods leads to superior BLEU and ROUGE performance over LC at low SNRs. 

\vspace{-1em}
\subsection{Impact of Code Length}
\label{section4d}
\vspace{-0.3em}
We compare four code configurations, including $(32,16)$, $(64,32)$, $(128,64)$, and $(256,128)$, at a fixed rate $R = 0.5$ to study the interplay between blocklength and semantic processing gain. Since the total sentence length is fixed at $512$ bits, varying the blocklength changes the number of segments to $q = 32, 16, 8, 4$, respectively.

\subsubsection{MSC-SEC Performance}
Figure~\ref{fig:sec} demonstrates BLER and semantic fidelity for MSC-SEC across all four code lengths, as well as their gaps to LC. As shown, shorter codes yield worse BLER at every SNR, consistent with trend indicated by finite blocklength bound. SEC can improve BLER for all code lengths (as shown in Fig. \ref{fig_overall}), but only the $(256,128)$ code reaches BLER comparable to LC under SEC. The remaining three codes, despite the SEC gain, still fall short of the LC waterfall curve by a considerable gap. The $(32,16)$ code performs worst, remaining above BLER $= 10^{-2}$ throughout the SNR range.

The semantic metrics exhibit the opposite trend. At low SNR, shorter codes achieve higher BLEU and ROUGE. Specifically, at $0$ dB, $(32,16)$ and $(64,32)$ lead $(128,64)$ and $(256,128)$ by roughly $24\%$ on both metrics. This is because each $(32,16)$ segment failure corrupts only $2$ characters, leaving most of the sentence intact for the language model to reconstruct from, whereas a $(256,128)$ failure corrupts $16$ characters and removes a larger fraction of the available context. As SNR increases and segment error rates drop, longer codes overtake shorter ones in semantic fidelity as well. This is because at high SNR the error-correction capability of longer channel codes dominates and the additional gain from semantic processing becomes marginal.

\subsubsection{MSC-SLD}

\begin{figure}[t]
    \centering
        \includegraphics[width=0.25\textwidth]{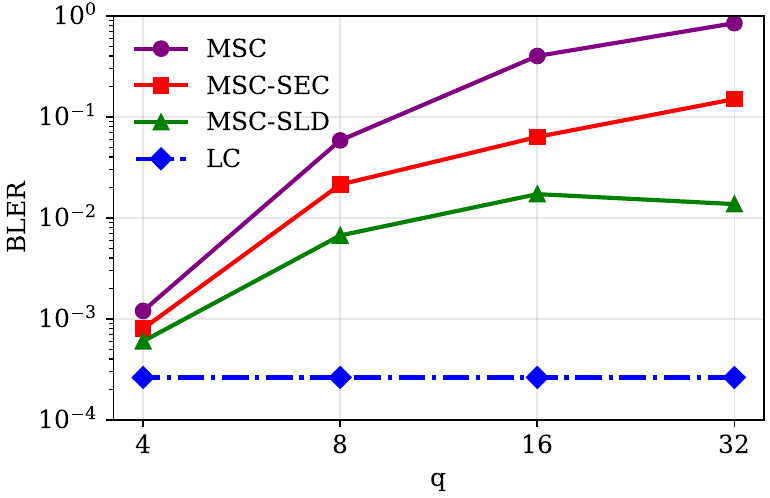}
    \vspace{-12pt}
    \caption{Impact of the number of segments $q$ on BLER at SNR $= 2$~dB for different methods.}
    \label{fig:impact_q}
    \vspace{-1.5em}
\end{figure}

Figure~\ref{fig:sld} shows MSC-SLD performance for all four code lengths. Compared to MSC-SEC in Fig.~\ref{fig:sec}, SLD provides additional BLER and semantic gains across all configurations, with the improvement most significant for shorter codes.

For BLER, at low SNR longer codes perform worse than shorter ones, because a longer code has fewer segments ($q = 4$ for $(256,128)$ versus $q = 32$ for $(32,16)$), leaving SLD with less intact context to reconstruct from. At high SNR, shorter codes exhibit higher BLER, as the channel code's own error-correction capability becomes the dominant factor and longer codes benefit from their larger minimum distance. The crossover occurs around $1.3$ dB. Among all configurations, the $(32,16)$ code reaches BLER $= 10^{-3}$ at $3$ dB, which is similar to the $(64,32)$ code.

For semantic fidelity, the advantage of shorter codes is more significant. At $0$ dB, the $(32,16)$ code achieves BLEU $= 94.3$ and ROUGE $= 97.4$, compared to $85$ and $92$ for $(64,32)$ and $61$ and $74$ for $(256,128)$. At the same SNR, LC produces near-zero BLEU and ROUGE because a single long-codeword failure destroys the entire sentence.

\subsubsection{BLER versus Number of Segments}
Figure~\ref{fig:impact_q} plots BLER against $q$ at a fixed SNR of $2$ dB. For MSC without semantic processing, BLER grows by nearly three orders of magnitude from $q = 4$ to $q = 32$, reflecting the finite-blocklength penalty of progressively shorter codes. Adding SEC and SLD narrows this spread. MSC-SLD limits the growth to approximately one order of magnitude over the same range. The gap between MSC and MSC-SLD widens with $q$, indicating that the semantic processing gain scales with the number of segments and partially compensates for the reduced coding gain of shorter blocklengths. At $q = 4$, MSC-SLD approaches the LC baseline.

\vspace{-1em}

\subsection{Analytical versus Simulation Results}
\begin{figure}[t]
    \centering
        \includegraphics[width=0.43\textwidth]{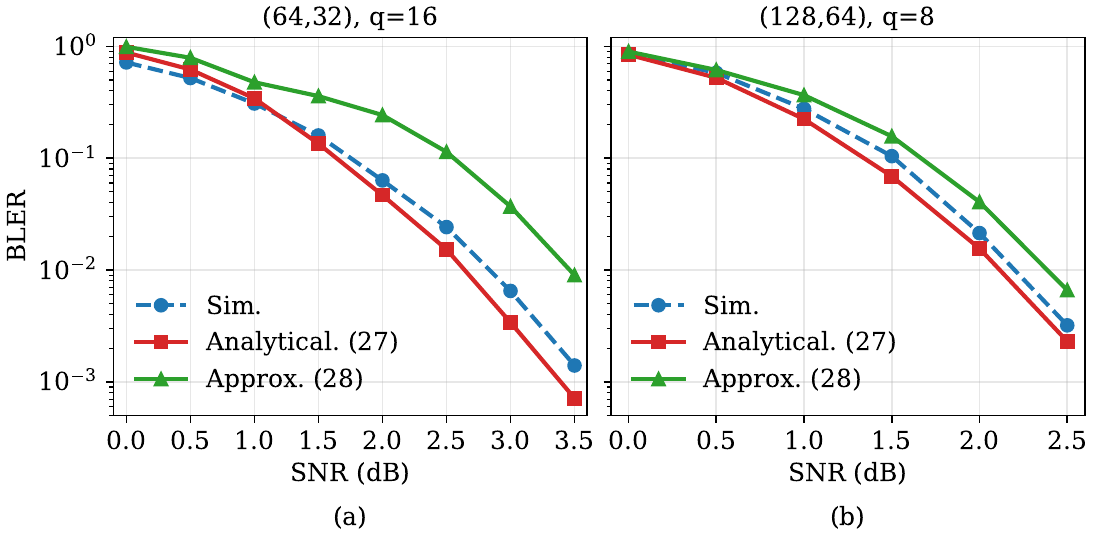}
    \vspace{-12pt}
    \caption{Comparison of analytical BLER and simulated BLER for MSC-SEC.}
    \label{fig:cal_vs_sim}
    \vspace{-8pt}
\end{figure}

\vspace{-0.3em}
Figure~\ref{fig:cal_vs_sim} compares the analytical BLER from (\ref{eq:cal_bler}) and the approximation from (\ref{eq:est_bler}) against simulation results for MSC-SEC using the $(64,32)$ and $(128,64)$ codes. The $(256,128)$ code is omitted from the figure as all three curves nearly overlap, confirming that both expressions are tight when the number of segments $q$ is small.

For the $(128,64)$ code with $q=8$, (\ref{eq:cal_bler}) closely matches the simulation, validating the independence assumption underlying the term $[P_{\text{rec}}(q_e)]^{q_e}$. As the code length decreases to $(64,32)$ with $q=16$, a noticeable gap emerges. Eq.~~(\ref{eq:cal_bler}) underestimates the simulated BLER by approximately 0.2~dB. This is because the independence assumption treats the recovery outcomes of individual erroneous segments as mutually independent, each succeeding with probability $P_{\text{rec}}(q_e)$. In practice, however, when multiple segments fail simultaneously, they mutually deprive each other of contextual information. The joint recovery probability is therefore lower than the product $[P_{\text{rec}}(q_e)]^{q_e}$, and this correlation effect becomes more pronounced as $q$ increases and simultaneous failures become more frequent.

The approximation (\ref{eq:est_bler}), which replaces $P_{\text{rec}}(q_e)$ with a constant $\eta$, overestimates the simulated BLER consistently. This is because $P_{\text{rec}}(q_e)$ decreases monotonically with $q_e$, and $\eta$, as a weighted average over all $q_e$, underestimates recovery for the small-$q_e$ terms that dominate the binomial sum at moderate $P_e^{\text{OSD}}$. Despite these gaps, both expressions track the simulated BLER trend closely across all code lengths and serve as practical bounds for code parameter selection.

\begin{figure}[t]
    \centering
        \includegraphics[width=0.27\textwidth]{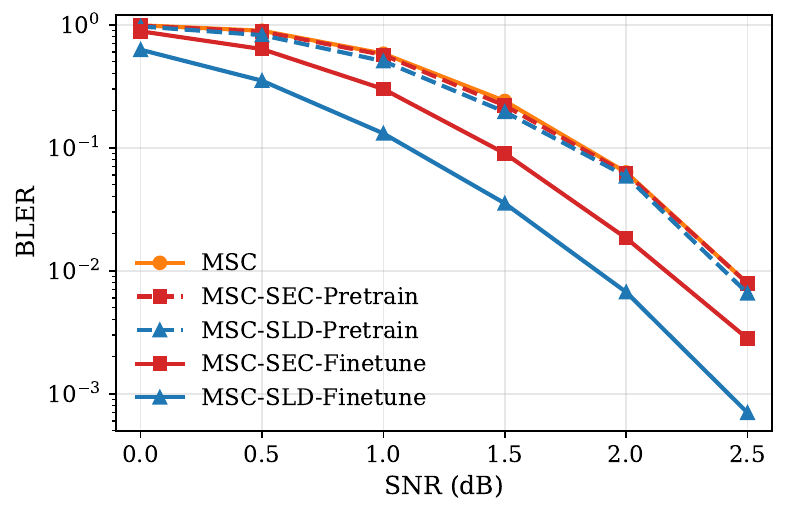}
    \vspace{-11pt}
    \caption{BLER comparison of pretrained and fine-tuned SEC/SLD models with $(128,64)$ MSC.}
    \label{fig:finetune_vs_pretrain}
    \vspace{-13pt}
\end{figure}

\vspace{-1em}
\subsection{Effect of Fine-Tuning}
\vspace{-0.3em}
The SEC and SLD modules rely on a fine-tuned BART model. A natural question is whether fine-tuning is necessary, or whether the pretrained BART-base already possesses sufficient language knowledge to perform the reconstruction. In this subsection, we replace the fine-tuned BART with the pretrained BART-base while keeping the rest of the pipeline. We test on the $(128,64)$ code.

As shown in Fig.~\ref{fig:finetune_vs_pretrain}, the pretrained models provide negligible improvement over the baseline MSC. This is because the generic denoising patterns learned during pretraining (e.g., token masking, deletion, and shuffling) differ substantially from the structured errors introduced by channel decoding. In contrast, the fine-tuned models achieve significant gains. SEC-Finetune provides approximately 0.4~dB improvement over MSC, and SLD-Finetune extends this to over 0.8~dB. These results confirm that fine-tuning on channel-specific error patterns is essential for the proposed framework.

\vspace{-1em}
\subsection{Retransmission Performance Analysis}
\vspace{-0.3em}
We evaluate HARQ performance with mother codes $\mathcal{C}(256, 64)$, $\mathcal{C}(128, 32)$, and $\mathcal{C}(64, 16)$ at mother-code rate $R_m = 1/4$ and first-transmission rate $R_t = 1/2$.
To ensure a fair comparison across configurations, the retransmission budget is fixed at $128$ parity bits per sentence across all three codes. This corresponds to retransmitting at most $1$ segment for $q = 8$, $2$ segments for $q = 16$, and $4$ segments for $q = 32$. Three schemes are compared:
(i) MSC-HARQ, which is the IR-HARQ baseline without semantic processing. It re-transmits segments randomly selected from the error set identified by $P_{\mathrm{SEC},i}$ in Section~\ref{section4a};
(ii) MSC-SLD-SHARQ, which applies SLD in both transmission rounds and prioritizes retransmission by the confidence score $P_{\mathrm{SLD},i}$. (iii) LC-HARQ, which encodes the entire sentence as a single long LDPC codeword and retransmits $128$ additional parity bits via IR-HARQ.

\begin{figure}[t]
    \centering
    \includegraphics[width=0.5\textwidth]{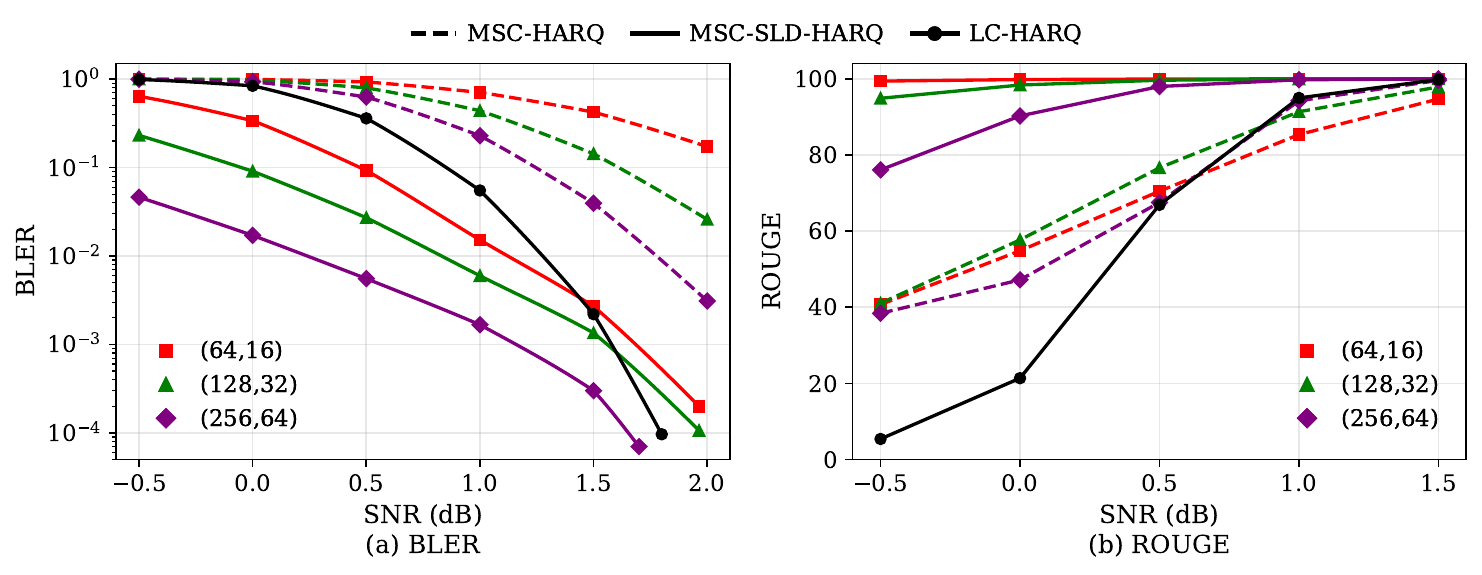}
    \vspace{-20pt}
    \caption{Retransmission performance comparison of MSC-HARQ, MSC-SLD-SHARQ, and LC-HARQ in BLER and ROUGE.}
    \label{fig:harq}
    \vspace{-1em}
\end{figure}

\subsubsection{BLER Performance} 
As shown in Fig.~\ref{fig:harq}~(a), MSC-HARQ suffers high BLER in the low-SNR regime, with the gap between code lengths widening as the SNR increases in accordance with the finite-blocklength bound. MSC-SLD-SHARQ achieves substantially lower BLER across the entire SNR range. The $(256, 64)$ code reaches a BLER of $0.046$ at $-0.5$ dB; MSC-HARQ requires approximately $2.1$ dB additional SNR to reach the same BLER. At high SNR, MSC-SLD-SHARQ with the $(256, 64)$ code retains a small advantage over LC-HARQ down to BLER~$= 10^{-4}$.

Under MSC-SLD-SHARQ, the BLER gap between $(128, 32)$ and $(64, 16)$ is much smaller than under MSC-HARQ. Shorter codes produce more segments, so retransmitting a few selected segments together with SLD recovery over the remaining context corrects a comparable fraction of errors. Both shorter codes maintain a BLER advantage over LC-HARQ up to 1.5~dB. Compared to single-transmission MSC-SLD as shown in Fig.~\ref{fig:sld}, MSC-SLD-SHARQ provides an additional gain of approximately 1.5~dB at the same target BLER.

\subsubsection{Semantic Performance} Figure~\ref{fig:harq}~(b) shows the ROUGE scores. Under MSC-HARQ, the three code lengths achieve similar ROUGE performance, with longer codes maintaining only a marginal lead. Although shorter codes suffer higher BLER due to weaker per-segment error correction, the retransmission resolves the most corrupted segments, and the remaining errors are distributed across many small segments, yielding comparable readability across code lengths.

Under MSC-SLD-SHARQ, shorter codes achieve markedly higher semantic scores. At $-0.5$ dB, the $(64, 16)$ code attains ROUGE $=\,99.4$, compared to $94.8$ for $(128, 32)$ and $76.1$ for $(256, 64)$, while all MSC-HARQ codes score approximately $40$ and LC-HARQ drops to $5.4$. Fine-grained segmentation localizes each failure to a few characters, so more context remains for language-model reconstruction than when a few large segments are missing. The $(64, 16)$ code maintains ROUGE~$>$~$99$ across all evaluated SNRs.

LC-HARQ exhibits highly unstable semantic performance. A successful LC decoding yields a perfect sentence, whereas a failure corrupts the entire output, leaving no partial context for semantic recovery. This produces near-zero ROUGE at low SNR even with retransmission.

\begin{figure}
    \centering
        \includegraphics[width=0.25\textwidth]{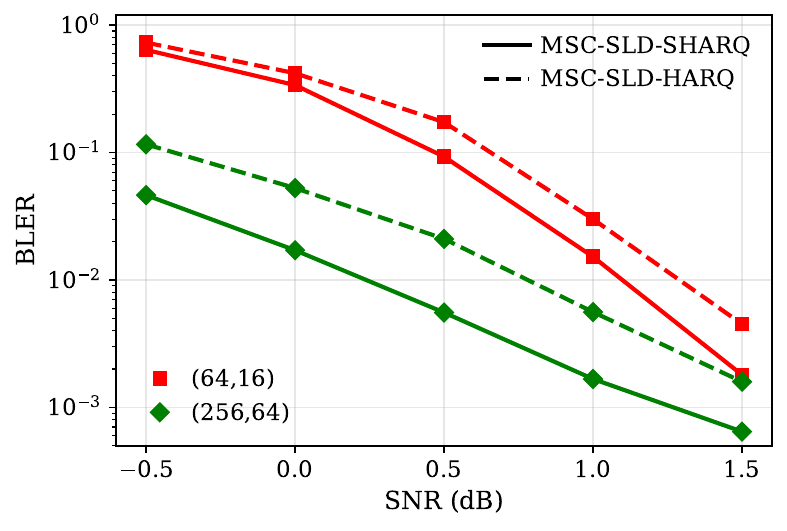}
    \vspace{-0.5em}
    \caption{Impact of retransmission strategy, confidence-guided (MSC-SLD-SHARQ) versus random selection (MSC-SLD-HARQ).}
    \label{fig:harq_vs_sharq}
    \vspace{-1em}
\end{figure}

\subsubsection{Retransmission Strategy Analysis} 
We compare two segment-selection strategies for retransmission: MSC-SLD-SHARQ, which retransmits the segments with the lowest success probability $P_{\mathrm{SLD},i}$, and its random-selection counterpart MSC-SLD-HARQ, which selects segments uniformly at random from the SLD error set $\mathcal{S}_{\mathrm{harq}}$. Both apply SLD in transmission rounds; they differ only in the retransmission priority.

As shown in Fig.~\ref{fig:harq_vs_sharq}, the confidence-guided strategy consistently outperforms random selection. MSC-SLD-SHARQ leads by $0.2$ dB for the $(64, 16)$ code and by $1.1$ dB for the $(256, 64)$ code. This trend reflects the distribution of $P_{\mathrm{SLD},i}$ across erroneous segments. For short codes with large $q$, per-segment BLER is high and the values of $P_{\mathrm{SLD},i}$ are relatively uniform, so random selection has a high probability of hitting a near-worst segment. For longer codes with small $q$, failures are few but their severity varies widely, so random selection is likely to miss the most damaging segment. The confidence-guided strategy targets the least reliable segment directly, yielding a larger marginal gain per retransmitted segment as $q$ decreases.

\vspace{-1em}
\subsection{Latency Analysis}
Table~\ref{tab:latency} presents the per-sentence decoding latency of each configuration. LC requires $1630$ ms per sentence with $80$ BP iterations, which is prohibitive for latency-sensitive applications. MSC can decode the $q$ segments in parallel. The per-sentence channel-decoding latency decreases with $q$, showing $160$, $90$, and $53$ ms for $q = 8$, $16$, and $32$, respectively. The growth with segment length is expected, since a longer segment requires a higher OSD order $m = \lceil d_{\min}/4 \rceil$, and the OSD complexity scales as $O(k^{m+2})$. The $(256,128)$ configuration ($q = 4$) is omitted because OSD at order $m = 8$ is computationally prohibitive. The semantic modules add a per-sentence cost that is independent of $q$, since each runs once on the full sentence, which are $63$ ms for SEC and $230$ ms for SLD.

Even with the semantic overhead included, MSC remains substantially faster than LC. At $q = 16$, MSC-SEC and MSC-SLD take $153$ ms and $383$ ms per sentence, corresponding to $90\%$ and $76\%$ latency reductions over LC, respectively. Combined with the BLER and semantic-fidelity results in Sections~\ref{section4c}, this confirms that the proposed framework improves reliability while retaining a latency profile suitable for short-packet transmission.

We note that BP decoding is itself amenable to parallelization, where all node updates can be executed concurrently under a flooding schedule. 
In this work we report the latency of a standard Sionna BP implementation.
The intra-decoder parallelism of BP applies to one long codeword, whereas MSC parallelism is segment-level. 
The segment-level parallelism remains complementary to the intra-decoder BP parallelism.

\begin{table}[t]
\centering
\caption{Latency breakdown for each decoding scheme (per sentence).}
\renewcommand{\arraystretch}{1}
\begin{tabular}{ll|ll}
\toprule
\textbf{Scheme} & \textbf{Time(ms)} & \textbf{Scheme} & \textbf{Time(ms)} \\
\midrule
LC & 1630 &  MSC ($q=8$) & 160\\
MSC ($q=32$) & 53 & MSC ($q=16$) & 90 \\
MSC-SEC & 63 & MSC-SLD & 230 \\
\bottomrule
\end{tabular}
\label{tab:latency}
\vspace{-1.5em}
\end{table}

\vspace{-1em}
\section{Conclusion}
\vspace{-0.5em}
This paper proposed a receiver-side framework that integrates a pretrained language model into the decoding of short block codes while preserving the source--channel separation architecture. The key insight is that transmitting a sentence as multiple short codewords localizes each channel-decoding failure to a small segment, leaving the surrounding segments to serve as context for language-model-based recovery. Building on this idea, the proposed semantic error correction (SEC), semantic list decoding (SLD), and semantic hybrid automatic repeat request (SHARQ) modules couple language-model inference with bit-level channel reliability through re-encoding, and replace CRC-based error detection with a joint semantic--channel confidence score. The resulting framework attains the BLER of a long-code baseline at a fraction of the decoding latency, while maintaining high semantic fidelity even in SNR regimes where the long code fails catastrophically.

\ifCLASSOPTIONcaptionsoff
\newpage
\fi
\vspace{-0.5em}
\bibliographystyle{IEEEtran}
\bibliography{IEEEabrv, refs}

@STRING{IEEE_J_SP         = "{IEEE} Trans. Signal Process."}

@STRING{IEEE_J_JSAC       = "{IEEE} J. Sel. Areas Commun."}

@STRING{IEEE_J_COM        = "{IEEE} Trans. Commun."}

@STRING{IEEE_J_WCOM       = "{IEEE} Trans. Wireless Commun."}

@STRING{IEEE_J_IT         = "{IEEE} Trans. Inf. Theory"}

@STRING{IEEE_J_CCN        = "{IEEE} Trans. on Cogn. Commun. Netw."}

@STRING{IEEE_M_COM        = "{IEEE} Commun. Mag."}

@misc{bit_flip,
  title={{Channel Coding for Unequal Error Protection in Digital Semantic Communication}},
  author={Kim, Seonjung and others},
  url={https://arxiv.org/pdf/2508.03381},
  archivePrefix={arXiv},
  year={2025},
}

@ARTICLE{hybird_semcom,
  author={Evgenidis, Nikos G. and others},
  journal=IEEE_J_WCOM, 
  title={Hybrid Semantic-Shannon Communications}, 
  year={2024},
  volume={23},
  number={9},
  pages={10926-10940},
  doi={10.1109/TWC.2024.3376998}}

@ARTICLE{pbosd,
  author={Yue, Chentao and others},
  journal={IEEE Commun. Lett.}, 
  title={Probability-Based Ordered-Statistics Decoding for Short Block Codes}, 
  year={2021},
  volume={25},
  number={6},
  pages={1791-1795},
  doi={10.1109/LCOMM.2021.3058978}}

@misc{beam_search,
  title={{Diverse Beam Search: Decoding Diverse Solutions from Neural Sequence Models}},
  author={Vijayakumar, Ashwin and others},
  url={https://arxiv.org/abs/1610.02424},
  archivePrefix={arXiv},
  year={2018},
}

@inproceedings{BPE,
    title = "Neural Machine Translation of Rare Words with Subword Units",
    author = "Sennrich, Rico and others",
    booktitle = "Proc. 54th Annu. Meeting Assoc. Comput. Linguist. (ACL)",
    year = "2016",
    pages = "1715--1725"
}

@ARTICLE{Lee,
  author={Lee, Ju-Hyung and others},
  journal=IEEE_J_WCOM , 
  title={Integrating Pre-Trained Language Model With Physical Layer Communications}, 
  year={2024},
  volume={23},
  number={11},
  pages={17266-17278},
  doi={10.1109/TWC.2024.3452481}}

@ARTICLE{9955525,
  author={Gündüz, Deniz and others},
  journal=IEEE_J_JSAC, 
  title={Beyond Transmitting Bits: Context, Semantics, and Task-Oriented Communications}, 
  year={2023},
  volume={41},
  number={1},
  pages={5-41}}

@ARTICLE{10494374,
  author={Gao, Shang and others},
  journal=IEEE_J_COM , 
  title={Importance of Semantic Information Based on Semantic Value}, 
  year={2024},
  volume={72},
  number={9},
  pages={5443-5457},
  doi={10.1109/TCOMM.2024.3385915}}

@ARTICLE{554278,
  author={Fossorier, M.P.C. and Shu Lin},
  journal=IEEE_J_COM , 
  title={Soft decision decoding of linear block codes based on ordered statistics for the Rayleigh fading channel with coherent detection}, 
  year={1997},
  volume={45},
  number={1},
  pages={12-14},
  doi={10.1109/26.554278}}

@ARTICLE{1057683,
  author={Gallager, R.},
  journal=IEEE_J_IT, 
  title={Low-density parity-check codes}, 
  year={1962},
  volume={8},
  number={1},
  pages={21-28},
  doi={10.1109/TIT.1962.1057683}}

@ARTICLE{6773024,
  author={Shannon, C. E.},
  journal={Bell Syst. Tech. J.}, 
  title={A mathematical theory of communication}, 
  year={1948},
  volume={27},
  number={3},
  pages={379-423},
  doi={10.1002/j.1538-7305.1948.tb01338.x}}

@inproceedings{bart,
    title = "{BART}: Denoising Sequence-to-Sequence Pre-training for Natural Language Generation, Translation, and Comprehension",
    author = "Lewis, Mike and others",
    booktitle = "Proc. Annu. Meeting Assoc. Comput. Linguistics (ACL)",
    month = jul,
    year = "2020",
    pages = "7871--7880"
}

@inproceedings{NIPS2017_3f5ee243,
 author = {Vaswani, Ashish and others},
 booktitle = {Proc. Adv. Neural Inf. Process. Syst. (NeurIPS)},
 pages = {},
 publisher = {Curran Associates, Inc.},
 title = {Attention is All You Need},
 volume = {30},
 year = {2017}
}

@misc{sionna,
  title={{Sionna: An open-source library for next-generation physical layer research}},
  author={J. Hoydis and others},
  url={https://arxiv.org/abs/2004.04913},
  archivePrefix={arXiv},
  year={2021},
}

@ARTICLE{PPV,
  author={Polyanskiy, Yury and Poor, H. Vincent and Verdu, Sergio},
  journal=IEEE_J_IT, 
  title={Channel Coding Rate in the Finite Blocklength Regime}, 
  year={2010},
  volume={56},
  number={5},
  pages={2307-2359},
  doi={10.1109/TIT.2010.2043769}}

@article{erseghe2016coding,
  title={Coding in the finite-blocklength regime: Bounds based on {L}aplace integrals and their asymptotic approximations},
  author={Erseghe, Tomaso},
  journal=IEEE_J_IT,
  volume={62},
  number={12},
  pages={6854--6883},
  year={2016},
  publisher={IEEE}
}

@ARTICLE{Mahyar2019ShortCode, 
author={M. {Shirvanimoghaddam} and others}, 
journal=IEEE_M_COM, 
title={Short Block-Length Codes for Ultra-Reliable Low Latency Communications}, 
year={2019}, 
volume={57}, 
number={2}, 
pages={130-137}, 
doi={10.1109/MCOM.2018.1800181}, 
ISSN={}, 
month={February},}

@article{yue2023efficient,
  title={Efficient decoders for short block length codes in {6G URLLC}},
  author={Yue, Chentao and others},
  journal=IEEE_M_COM,
  volume={61},
  number={4},
  pages={84--90},
  year={2023},
  publisher={IEEE}
}

@INPROCEEDINGS{jscc,
  author={Bourtsoulatze and others},
  booktitle={Proc. IEEE Int. Conf. Acoust., Speech Signal Process. (ICASSP)}, 
  title={Deep Joint Source-channel Coding for Wireless Image Transmission}, 
  year={2019},
  volume={},
  number={},
  pages={4774-4778},
  doi={10.1109/ICASSP.2019.8683463}}

@ARTICLE{deepsc,
  author={Xie, Huiqiang and others},
  journal=IEEE_J_SP, 
  title={Deep Learning Enabled Semantic Communication Systems}, 
  year={2021},
  volume={69},
  number={},
  pages={2663-2675}}

@ARTICLE{d2jscc,
  author={Huang, Jianhao and others},
  journal=IEEE_J_JSAC, 
  title={D²-JSCC: Digital Deep Joint Source-Channel Coding for Semantic Communications}, 
  year={2025},
  volume={43},
  number={4},
  pages={1246-1261},
  doi={10.1109/JSAC.2025.3531546}}

@ARTICLE{swin-jscc,
  author={Yang, Ke and others},
  journal=IEEE_J_CCN, 
  title={SwinJSCC: Taming Swin Transformer for Deep Joint Source-Channel Coding}, 
  year={2025},
  volume={11},
  number={1},
  pages={90-104},
  doi={10.1109/TCCN.2024.3424842}}

@article{LIU,
title = {Extended context-based semantic communication system for text transmission},
journal = {Digit. Commun. Netw.},
volume = {10},
number = {3},
pages = {568-576},
year = {2024},
issn = {2352-8648},
author = {Yueling Liu and others},
}

@inproceedings{snli,
    title = "A large annotated corpus for learning natural language inference",
    author = "Bowman, Samuel R. and others",
    booktitle = "Proc. Conf. Empirical Methods Natural Language Process. (EMNLP)",
    month = sep,
    year = "2015",
    pages = "632--642"
}

@inproceedings{bleu,
    title = "{B}leu: a Method for Automatic Evaluation of Machine Translation",
    author = "Papineni, Kishore and others",
    booktitle = "Proc. Annu. Meeting Assoc. Comput. Linguistics (ACL)",
    month = jul,
    year = "2002",
    pages = "311--318"
}

@inproceedings{rouge,
    title = "{ROUGE}: A Package for Automatic Evaluation of Summaries",
    author = "Lin, Chin-Yew",
    booktitle = "Text Summarization Branches Out",
    month = jul,
    year = "2004",
    pages = "74--81"
}

@INPROCEEDINGS{secrecy,
  author={Mu, Xidong and Liu, Yuanwei},
  booktitle={Proc. IEEE Int. Conf. Commun. (ICC)}, 
  title={Semantic Communication-Assisted Physical Layer Security Over Fading Wiretap Channels}, 
  year={2024},
  volume={},
  number={},
  pages={2101-2106},
  doi={10.1109/ICC51166.2024.10622708}}

@ARTICLE{MIMO1,
  author={Wu, Minghui and others},
  journal=IEEE_J_JSAC, 
  title={Deep Joint Semantic Coding and Beamforming for Near-Space Airship-Borne Massive MIMO Network}, 
  year={2025},
  volume={43},
  number={1},
  pages={260-278},
  doi={10.1109/JSAC.2024.3460084}}

@misc{TWRC,
  title={{Semantic Communication-Empowered Physical-layer Network Coding}},
  author={S. Yang and others},
  url={https://arxiv.org/abs/2209.00791},
  archivePrefix={arXiv},
  year={2022},
}

@misc{LLM-aid,
  title={{Semantic Pilot Design for Data-Aided Channel Estimation Using a Large Language Model}},
  author={S. Park and H. Yang},
  url={https://arxiv.org/abs/2602.04126},
  archivePrefix={arXiv},
  year={2025},
}

@misc{harq,
  title={{Semantic HARQ for Intelligent Transportation Systems: Joint Source-Channel Coding-Powered Reliable Retransmissions}},
  author={Y. Li and others},
  url={https://arxiv.org/abs/2504.14615},
  archivePrefix={arXiv},
  year={2025},
}

@ARTICLE{yue2025guesswork,
  author={Yue, Chentao and She, Changyang and Vucetic, Branka and Li, Yonghui},
  journal=IEEE_J_IT, 
  title={The Guesswork of Ordered Statistics Decoding: Guesswork Complexity and Decoder Design}, 
  year={2025},
  volume={71},
  number={6},
  pages={4167-4192},
  doi={10.1109/TIT.2025.3559168}}

@manual{etsi38212,
  title        = {{ETSI TS 138 212 V16.2.0: 5G; NR; Multiplexing and Channel Coding}},
  organization = {European Telecommunications Standards Institute (ETSI)},
  year         = {2020}
}
\vspace{-1em}

\end{document}